\documentclass[aps,prl,floatfix,twocolumn,superscriptaddress]{revtex4-2}

\usepackage{graphicx,hyperref,siunitx,titlesec,bm}
\usepackage{subfigure}
\usepackage[toc,page]{appendix}
\usepackage{float}

\hypersetup{
      colorlinks=true,
      citecolor=blue,
      linkcolor=blue,
      urlcolor=blue}
      
\usepackage{amsmath}
\usepackage{amsfonts}
\usepackage{amssymb}
\usepackage{braket}
\usepackage{bm}
\usepackage{times}
\usepackage{scalerel}

\usepackage{tikz}
\usetikzlibrary{svg.path}

\usepackage{upgreek}
\usepackage{algpseudocode}
\usepackage{algorithm}

\graphicspath{{figures/}}
\newcommand\wordcount{\input{|"texcount -inc -sum -0 -template={SUM} \jobname.tex"}}

\newcommand{\myref}[2][]{Fig.~\hyperref[#2]{\ref*{#2}#1}}
\newcommand{\Myref}[2][]{Figure~\hyperref[#2]{\ref*{#2}#1}}

\newcommand{\klaus}{\cite{Klaus2022oov}}

\newcommand{\edd}{\epsilon_{dd}}

\definecolor{orcidlogocol}{HTML}{A6CE39}
\tikzset{
  orcidlogo/.pic={
    \fill[orcidlogocol] svg{M256,128c0,70.7-57.3,128-128,128C57.3,256,0,198.7,0,128C0,57.3,57.3,0,128,0C198.7,0,256,57.3,256,128z};
    \fill[white] svg{M86.3,186.2H70.9V79.1h15.4v48.4V186.2z}
                 svg{M108.9,79.1h41.6c39.6,0,57,28.3,57,53.6c0,27.5-21.5,53.6-56.8,53.6h-41.8V79.1z M124.3,172.4h24.5c34.9,0,42.9-26.5,42.9-39.7c0-21.5-13.7-39.7-43.7-39.7h-23.7V172.4z}
                 svg{M88.7,56.8c0,5.5-4.5,10.1-10.1,10.1c-5.6,0-10.1-4.6-10.1-10.1c0-5.6,4.5-10.1,10.1-10.1C84.2,46.7,88.7,51.3,88.7,56.8z};
  }
}

\newcommand\orcidicon[1]{\href{https://orcid.org/#1}{\mbox{\scalerel*{
\begin{tikzpicture}[yscale=-1,transform shape]
\pic{orcidlogo};
\end{tikzpicture}
}{|}}}}

\setcitestyle{super}

\begin{document}
\title{Observation of vortices in a dipolar supersolid}

\author{Eva Casotti\,\orcidicon{0000-0002-8340-1445}}
\thanks{These authors contributed equally to this work.}
\affiliation{Institut f\"{u}r Quantenoptik und Quanteninformation, \"Osterreichische Akademie der \\ Wissenschaften, Technikerstr. 21A, 6020 Innsbruck, Austria }
\affiliation{Universität Innsbruck, Fakultät für Mathematik, Informatik und Physik,
Institut für Experimentalphysik, 6020 Innsbruck, Austria}

\author{Elena Poli\,\orcidicon{0000-0002-1295-9097}}
\thanks{These authors contributed equally to this work.}
\affiliation{Universität Innsbruck, Fakultät für Mathematik, Informatik und Physik,
Institut für Experimentalphysik, 6020 Innsbruck, Austria}

\author{Lauritz Klaus\,\orcidicon{0000-0002-6018-0811}}
\affiliation{Institut f\"{u}r Quantenoptik und Quanteninformation, \"Osterreichische Akademie der \\ Wissenschaften, Technikerstr. 21A, 6020 Innsbruck, Austria }
\affiliation{Universität Innsbruck, Fakultät für Mathematik, Informatik und Physik,
Institut für Experimentalphysik, 6020 Innsbruck, Austria}

\author{Andrea Litvinov\,\orcidicon{0009-0001-5332-1188}}
\affiliation{Institut f\"{u}r Quantenoptik und Quanteninformation, \"Osterreichische Akademie der \\ Wissenschaften, Technikerstr. 21A, 6020 Innsbruck, Austria }

\author{Clemens Ulm\,\orcidicon{0009-0009-7589-9536}}
\affiliation{Institut f\"{u}r Quantenoptik und Quanteninformation, \"Osterreichische Akademie der \\ Wissenschaften, Technikerstr. 21A, 6020 Innsbruck, Austria }
\affiliation{Universität Innsbruck, Fakultät für Mathematik, Informatik und Physik,
Institut für Experimentalphysik, 6020 Innsbruck, Austria}

\author{Claudia Politi\,\orcidicon{0000-0001-7236-7926}}
\thanks{Current address: Institute for Quantum Electronics, ETH Zürich, Otto-Stern-Weg 1, 8093 Zürich, Switzerland}
\affiliation{Institut f\"{u}r Quantenoptik und Quanteninformation, \"Osterreichische Akademie der \\ Wissenschaften, Technikerstr. 21A, 6020 Innsbruck, Austria }
\affiliation{Universität Innsbruck, Fakultät für Mathematik, Informatik und Physik,
Institut für Experimentalphysik, 6020 Innsbruck, Austria}

\author{Manfred J. Mark\,\orcidicon{0000-0001-8157-4716}}
\affiliation{Universität Innsbruck, Fakultät für Mathematik, Informatik und Physik,
Institut für Experimentalphysik, 6020 Innsbruck, Austria}
\affiliation{Institut f\"{u}r Quantenoptik und Quanteninformation, \"Osterreichische Akademie der \\ Wissenschaften, Technikerstr. 21A, 6020 Innsbruck, Austria }

\author{Thomas Bland\,\orcidicon{0000-0001-9852-0183}}
\affiliation{Universität Innsbruck, Fakultät für Mathematik, Informatik und Physik,
Institut für Experimentalphysik, 6020 Innsbruck, Austria}

\author{Francesca Ferlaino\,\orcidicon{0000-0002-3020-6291}}
\thanks{Correspondence should be addressed to: \mbox{\url{francesca.ferlaino@uibk.ac.at}}}
\affiliation{Universität Innsbruck, Fakultät für Mathematik, Informatik und Physik,
Institut für Experimentalphysik, 6020 Innsbruck, Austria}
\affiliation{Institut f\"{u}r Quantenoptik und Quanteninformation, \"Osterreichische Akademie der \\ Wissenschaften, Technikerstr. 21A, 6020 Innsbruck, Austria }

\date{\today}

\begin{abstract} 
Supersolids are states of matter that spontaneously break two continuous symmetries: translational invariance due to the appearance of a crystal structure and phase invariance due to phase locking of single-particle wave functions, responsible for superfluid phenomena. While originally predicted to be present in solid helium\cite{Gross1957uto,Gross1958cto,Andreev1969qto,Chester1970sob,Leggett1970cas}, ultracold quantum gases provided a first platform to observe supersolids\cite{Li2017asp,Leonard2017sfi,Boettcher2019tsp,Tanzi2019ooa,Chomaz2019lla}, with particular success coming from dipolar atoms\cite{Boettcher2019tsp,Tanzi2019ooa,Chomaz2019lla,Norcia2021tds,Chomaz2022dpa}. Phase locking in dipolar supersolids has been probed through e.g.~measurements of the phase coherence\cite{Boettcher2019tsp,Tanzi2019ooa,Chomaz2019lla} and gapless Goldstone modes\cite{Guo2019tle}, but quantized vortices, a hydrodynamic fingerprint of superfluidity, have not yet been observed. Here, with the prerequisite pieces at our disposal, namely a method to generate vortices in dipolar gases\cite{Klaus2022oov,Bland2023vid} and supersolids with two-dimensional crystalline order\cite{Norcia2021tds,Norcia2022cao,Bland2022tds}, we report on the theoretical investigation and experimental observation of vortices in the supersolid phase. Our work reveals a fundamental difference in vortex seeding dynamics between unmodulated and modulated quantum fluids. This opens the door to study the hydrodynamic properties of exotic quantum systems with multiple spontaneously broken symmetries, in disparate domains such as quantum crystals and neutron stars\cite{Poli2023gir}. 
\end{abstract} 

\maketitle

\noindent Rotating fluids on all scales exhibit a whirling motion known as vorticity. Unique to the quantum world, however, is the quantization of this rotation due to the single-valued and continuous nature of the underlying macroscopic wavefunction\cite{Onsager1949ido,Feynman1955cia}. Observing quantized vortices is regarded as unambiguous evidence of superfluidity, relevant for a wide variety of interacting many-body quantum systems from superfluid $^4$He\,\cite{Yarmchuk1979oos,Bewley2006voq} through gaseous bosonic\cite{AboShaeer2001oov} and fermionic\cite{Zwierlein2005vas} condensates, exciton-polariton condensates\cite{Lagoudakis2008qvi}, to solid-state type-II superconductors\cite{Wells2015aol,Embon2017ios}. Remarkably, this phenomenon persists over a wide range of interaction scales, since it requires only the irrotational nature of the velocity field. All of these examples refer to the case in which there is a single spontaneously broken symmetry, leading to the question: what new properties do we expect to arise in systems with multiple broken symmetries?

The supersolid phase belongs to this category, spontaneously breaking phase and translational symmetries. Supersolids, characterized by the coexistence of superfluid and solid properties\cite{Gross1957uto,Gross1958cto,Andreev1969qto,Chester1970sob,Leggett1970cas}, have been investigated through two distinct approaches. Either imbuing superfluid properties into a solid\cite{Hamidian2016doa, Nyeki2017isa, Levitin2019efa, Agterberg2020tpo, Liu2023pdw}, or partially crystallizing a superfluid\cite{Leonard2017sfi, Li2017asp,Boettcher2019tsp,Tanzi2019ooa,Chomaz2019lla,Norcia2021tds}.
Among these systems, supersolids composed of dipolar atoms have emerged as a versatile platform for exploring the superfluid characteristics and solid properties of this long sought-after state\cite{Chomaz2022dpa}. Where these studies have found a roadblock is in probing the response to rotation. One consequence of irrotational flow is the scissors mode oscillation, where the signature of superfluidity is the lack of a rigid body response to a sudden rotation of an anisotropic trap\cite{Marago2000oot}. However, supersolids show a mixture of rotational and irrotational behavior, leading to a multimode response to perturbation. This complexity hinders a straightforward extraction of the superfluid contribution\cite{Tanzi2021eos,Norcia2022cao,Roccuzzo2022moi}. Instead, the presence of quantized vortices is an unequivocal signal of irrotationality, and thus unambiguously proves the superfluidity of the system. These vortices are also anticipated to exhibit other distinctive characteristics, including a reduced angular momentum\cite{Gallemi2020qvi,Roccuzzo2020ras}, and unusual dynamics due to their interplay with the crystal such as pinning and snaking\cite{Henkel2012svc,Ancilotto2021vpi,Poli2023gir}. Investigating these dynamics could provide new insights into flux pinning in superconductors\cite{Matsushita2014fpi} and glitches in neutron stars\cite{Poli2023gir}. Nevertheless, a critical gap exists in the current experimental exploration of supersolids — an investigation into whether the supersolid can maintain its structure and coherence under continuous stirring, as well as if, and how, vortices may manifest and behave in this unique state. The experimental challenge lies in the inherent complexity and fragility of the supersolid phase, which lives in a narrow region within the phase diagram\cite{Chomaz2022dpa}.
In our work, we explore this uncharted territory by investigating the supersolid response to rotation, using a technique known as magnetostirring\cite{Prasad2019vlf, Klaus2022oov, Bland2023vid}. Combining experiment and theory, our study explores both the unmodulated and modulated states, revealing distinctive signatures associated with the presence of vortices in the supersolid.

\vspace{0.1cm}
\noindent {\bf Predicting the supersolid response to rotation} 

\noindent Owing to the inherent long-range interactions among atoms, a dipolar gas exhibits a density distribution that extends along the magnetic field direction, a phenomenon known as magnetostriction\cite{Stuhler2005ood}. This imparts an elliptical shape to the cloud. The rotation of the magnetic field consequently induces stirring of the gas\cite{Prasad2019vlf}. This method, referred to as magnetostirring, has recently been employed to generate vortices in unmodulated dipolar quantum gases\cite{Klaus2022oov}. These vortices eventually organize into distinctive patterns, forming either triangular or stripe vortex lattices\cite{Prasad2019vlf,Bland2023vid}. 

Generating vortices in the supersolid phase through magnetostirring has not yet been investigated, therefore, we theoretically explore the zero temperature dynamics of our state through the so-called extended Gross--Pitaevskii equation\cite{Waechtler2016qfi,FerrierBarbut2016ooq,Chomaz2016qfd,Bisset2016gsp} (eGPE). This takes into account the cylindrically symmetric harmonic trap, the short-range interactions, through the tunable s-wave interaction strength $a_s$, and long-range interactions, with fixed amplitude $a_\text{dd}\,{=}\,130.8a_0$ for $^{164}$Dy. Also included are beyond-mean field effects resulting from the zero-point energy of Bogoliubov quasiparticles--shown to be crucial for the energetic stability in the supersolid phase\cite{Bisset2016gsp}. By tuning the short-range interactions, we can access both the supersolid (typically $\epsilon_\text{dd}\,{=}\,a_\text{dd}/a_s\,{\gtrsim}\,1.3$ for experimentally relevant trap geometries) and unmodulated Bose--Einstein condensate (BEC) phases ($\epsilon_\text{dd}\,{\lesssim}\,1.3$).

\begin{figure}[t]
\centering    
\includegraphics[width=0.9\columnwidth]{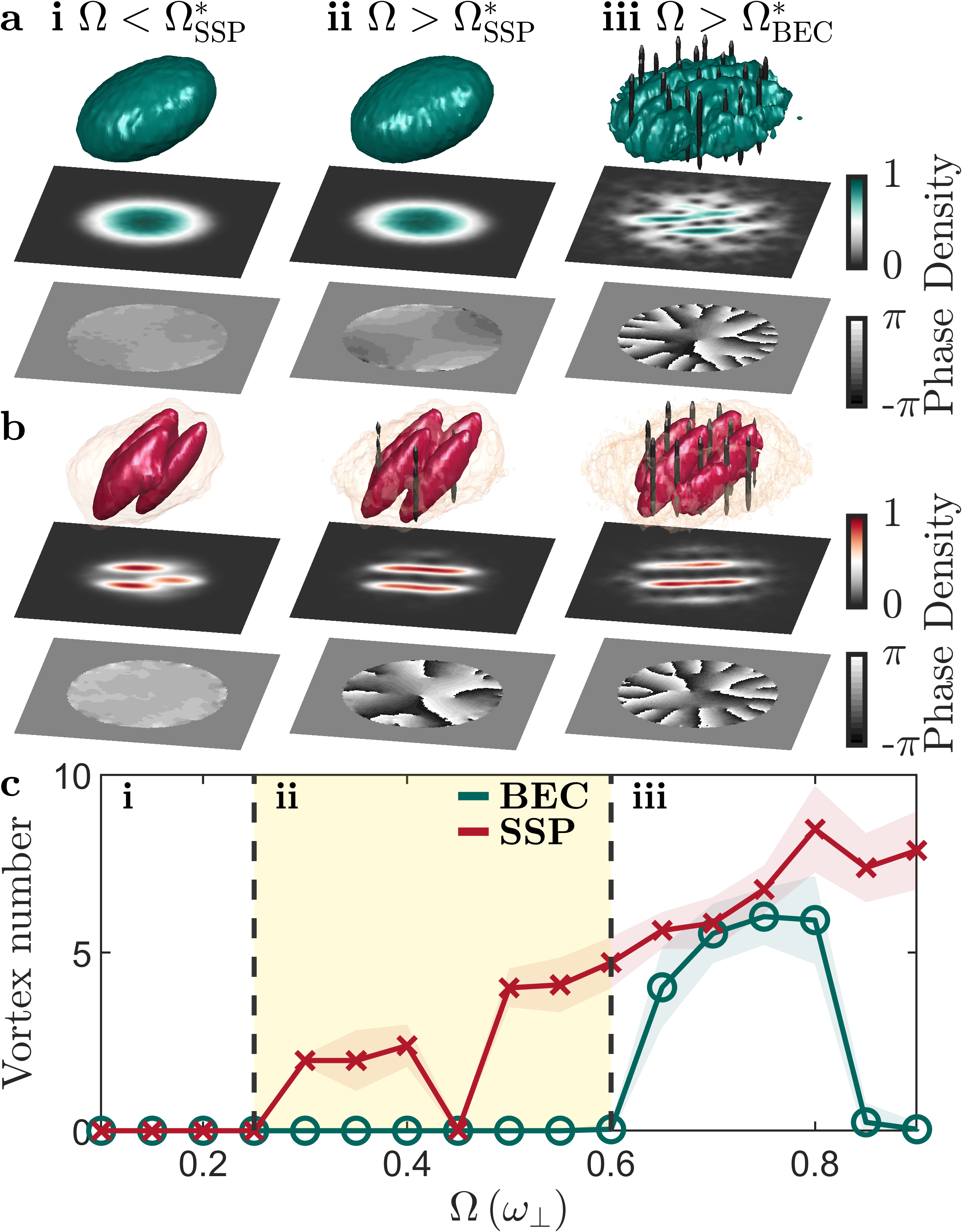}
\vspace{-0.3cm}
    \caption{\textbf{Simulation of vortex nucleation in a supersolid and unmodulated BEC.} Density isosurfaces and their corresponding normalized integrated density and phase profiles for the \textbf{a} unmodulated BEC and \textbf{b} supersolid phases after $1\,\si{s}$ of rotation at (i) $\Omega\,{=}\,0.2\omega_\perp$, (ii) $0.4\omega_\perp$, and (iii) $0.7\omega_\perp$. Isosurfaces are shown at $15\%$ of the max density in all plots, and additionally at $0.5\%$ in the SSP to show the halo. Vortex tubes are shown in black in the 3D images and appear as $2\pi$ windings in the phase plots. \textbf{c} Comparison of the time-averaged vortex number as a function of $\Omega$ between the SSP (red) and BEC (green), averaged between $0.75\,\si{s}$ and $1\,\si{s}$ of rotation, and the colored shading shows the standard deviation. The yellow shaded area highlights $\Omega^*_\mathrm{SSP}<\Omega<\Omega^*_\mathrm{BEC}$ (see main text). The results are obtained from eGPE calculations with $(\omega_\perp,\,\omega_z)\,{=}\,2\pi\,{\times}\,[50,103]\,\si{Hz}$, magnetic-field angle from the $z$-axis $\theta\,{=}\,30^\circ$, atom number $N\,{=}\,5\,{\times}\,10^4$, and scattering length $a_s\,{=}\,95a_0$ ($104a_0$) for the SSP (BEC) phase.}
    \label{fig:Fig1}
\end{figure}

Figure \ref{fig:Fig1} comparatively shows exemplar density and phase distributions of an unmodulated dipolar BEC [a] and supersolid phase (SSP) [b] rotating the magnetic field at increasing frequency $\Omega$, from left to right. In a BEC, Fig.\,\ref{fig:Fig1}a, at small frequencies with respect to the radial trap frequency $\omega_\perp$, the cloud density is almost unchanged from the static result [a(i)]. Rotating faster, the cloud elongates, and we observe an irrotational velocity field in the phase profile [a(ii)]. When rotating faster than a given $\Omega^*_\text{BEC}$, the irrotational flow can no longer be maintained, and quantum vortices, observable as density holes and quantized $2\pi$ phase windings, penetrate the condensate surface following a quadrupole mode instability [a(iii)]\cite{Klaus2022oov}.

In contrast to unmodulated BECs, supersolids present a new scenario, see Fig.\,\ref{fig:Fig1}b. Our simulations reveal that the system is more susceptible to quantized vortex creation, happening at significantly lower frequencies than the BEC case. At small frequencies, the crystalline structure and surrounding `halo' of atoms follow the magnetic field in lockstep without generating vortices [b(i)]. At higher frequencies, yet still $\Omega<\Omega^*_\text{BEC}$, we now see vortex lines smoothly entering into the interstitial regions between the crystal sites [b(ii)]\cite{Gallemi2020qvi,Roccuzzo2020ras,Ancilotto2021vpi}. These vortices persist even at higher frequencies, arranging into a regular lattice structure [b(iii)].

To gain further insight, we study the total vortex number obtained after $1\,\si{s}$ of rotation as a function of $\Omega$. Figure\,\ref{fig:Fig1}c shows a striking difference in the response to rotation between the two quantum phases. The BEC shows the well-known resonant behavior, in which the rotation frequency must be at resonance with half the collective quadrupole mode frequency $\omega_Q$. This drives an instability of the condensate surface, allowing vortices to enter the state. For a non-dipolar BEC $\omega_Q\,{=}\,\omega_\perp/\sqrt{2}$\cite{Recati2001oco,Sinha2001dio,Madison2001sso}, while for dipolar quantum gases, small deviations from this value can occur depending on the dipolar interaction and the trap geometry\cite{vanBijnen2007dio}. For our system, we see the onset of the resonant behavior at $\Omega^*_\text{BEC}\,{=}\,0.6\omega_\perp$, reaching its maximum at $\Omega\approx0.75\omega_\perp$.

In the supersolid phase, we observe a vastly different behavior. The dual superfluid-crystalline nature of the state leads to two distinguishing features: the reduced superfluidity results in vortices nucleating at a lower rotation frequency and the solidity gives rise to a monotonic increase in vortex number at faster frequencies, reminiscent of rigid body rotation; see Fig.\,\ref{fig:Fig1}c. This can be understood by studying the excitation spectrum. A two-dimensional supersolid exhibits three quadrupole modes: one from the broken phase symmetry associated with superfluidity and one from each direction of the broken translational symmetry\cite{Gallemi2020qvi}. In our case, the latter are nearly degenerate due to the cylindrically symmetric dipole trap. Excitation of the `superfluid' quadrupole mode is responsible for the weak resonance starting at $\Omega^*_\text{SSP}\,{\approx}\,0.25\omega_\perp$ and centered around $\Omega\,{\approx}\,0.35\omega_\perp$, where just a few vortices are created. The position of this resonance is highly dependent on the superfluid fraction, eventually vanishing in the so-called isolated droplet (ID) regime\cite{Gallemi2020qvi}. In the ID regime, there is no phase coherence nor density between the droplets, and therefore the vortex number is zero. A state that is initially in the supersolid phase cannot be rotated into the ID regime. This state is discussed theoretically in more detail in the \nameref{appendix:Methods}. However, we note that it is out of current experimental reach to create an ID state with a lifetime on the order of the vortex seeding time. As we will discuss later, the detection of the low frequency resonance is at the edge of our current experimental capability, indicating compatibility, albeit with a low signal strength. Beginning at $\Omega\,{\approx}\,0.45\omega_\perp$, the system exhibits instead a threshold response to rotation, where the angular momentum, and thus vortex number, linearly increases with $\Omega$\cite{Roccuzzo2020ras,Gallemi2020qvi}. This prominent feature arises due to the near degenerate crystal quadrupole mode resonance.

\vspace{0.1cm}
\noindent {\bf Experimental magnetostirring of a dipolar supersolid}

\noindent Bolstered by the acquired theoretical understanding, we experimentally explore the suitability of magnetostirring to nucleate vortices in the supersolid phase. We first produce an optically trapped supersolid quantum gas of highly magnetic bosonic $^{164}$Dy atoms via direct evaporative cooling\cite{Chomaz2019lla, Norcia2021tds,Sohmen2021bla,Bland2022tds} and then apply magnetostirring\cite{Prasad2019vlf,Klaus2022oov,Bland2023vid} to rotate the gas.

\begin{figure}[t]
\includegraphics[width=1\columnwidth]{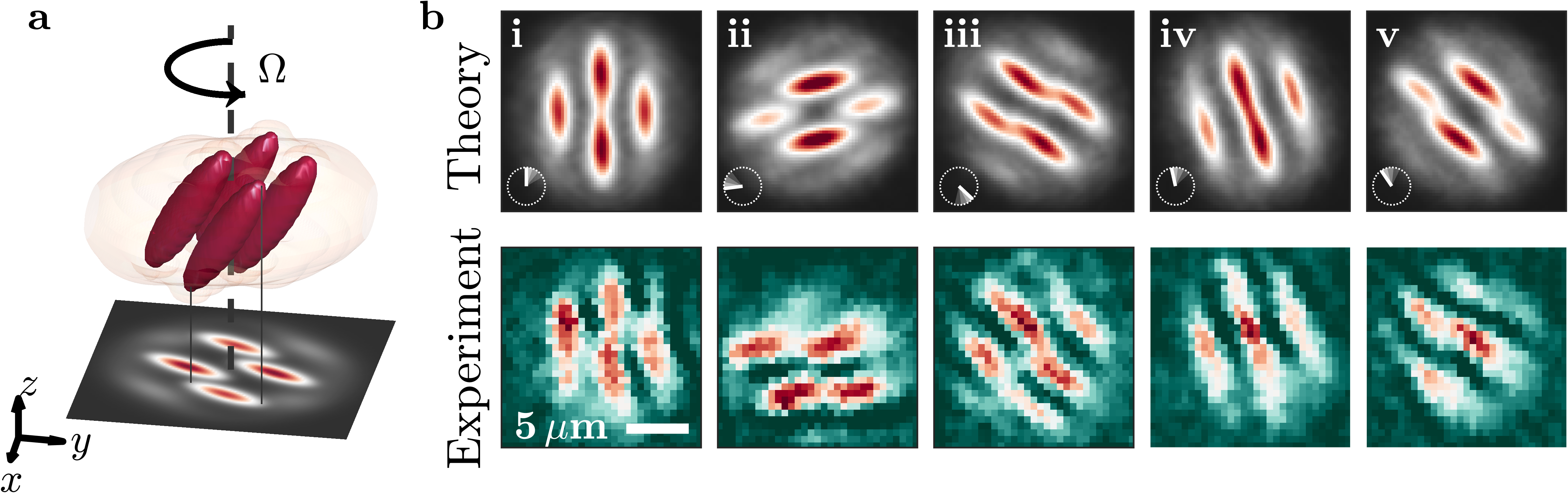}
\caption{\textbf{Magnetostirring of a $^{164}$Dy dipolar supersolid.} \textbf{a} Density isosurfaces shown at $15\%$ and $0.5\%$ of the maximum density and corresponding integrated density of a four droplet supersolid. \textbf{b} Column densities of a four droplet supersolid state from theory (top row) and experiment (bottom row) with $\Omega\,{=}\,0.3\omega_\perp$; the images were taken after (i-v) $1, 19, 43, 70, 274\,\si{ms}$. The insets show the rotation of the magnetic field vector in the $x$-$y$ plane with white lines. Experimental parameters: $B\,{=}\,18.24(2)\,\si{G}$, $N\,{\approx}\,7\,{\times}\,10^4$, and $(\omega_\perp,\omega_z)\,{=}\,2\pi\,{\times}\,[50.5(3),135(2)]\,\si{Hz}$. Illustrative simulation parameters: $a_s\,{=}\,92.5a_0$, $N\,{=}\,6\,{\times}\,10^4$, and $(\omega_\perp,\omega_z)\,{=}\,2\pi\,{\times}\,[50,135]\,\si{Hz}$.}
\label{fig:Fig2}
\end{figure}

In all the experiments presented, the three-dimensional optical dipole trap (ODT) is cylindrically symmetric, with radial frequency $\omega_\perp\,{\approx}\,2\pi\,{\times}\,50\,\si{Hz}$ and a trap aspect ratio $\omega_z/\omega_\perp$ that varies between $2$ and $3$. Throughout the evaporation sequence, we apply a uniform magnetic field along the $z$-axis and tilt the magnetic field vector by $\theta\,{=}\,30^\circ$ in the last cooling stage to prepare for magnetostirring\cite{Klaus2022oov}. With this sequence, we obtain a supersolid typically composed of four density maxima (droplets) on top of a low-density background (halo) of coherent atoms, which we verify with a measurement of the phase coherence after long time-of-flight (see \nameref{appendix:Methods}). Taking phase-contrast images after $3\,\si{ms}$ of expansion gives us access to the 2D density profiles integrated along the axial direction, as illustrated in Fig.\,\ref{fig:Fig2}a. We magnetostir the system by rotating the magnetic-field vector around the $z$-axis with a constant angular velocity $\Omega$; see Fig.\,\ref{fig:Fig2}b. As predicted by theory, the droplets align themselves along the magnetic-field direction, breaking the cylindrical symmetry, thus enabling rotation. We are able to stir the supersolid for hundreds of milliseconds without destroying the state, as shown in Fig.\,\ref{fig:Fig2}b(i-v). This result is particularly relevant since it allows several full rotations, even for small driving frequencies, giving the vortices enough time to nucleate and percolate into the system.

\vspace{0.1cm} 
\noindent {\bf Observation of vortices in a dipolar supersolid}

\begin{figure*}[t]
\includegraphics[width=2\columnwidth]{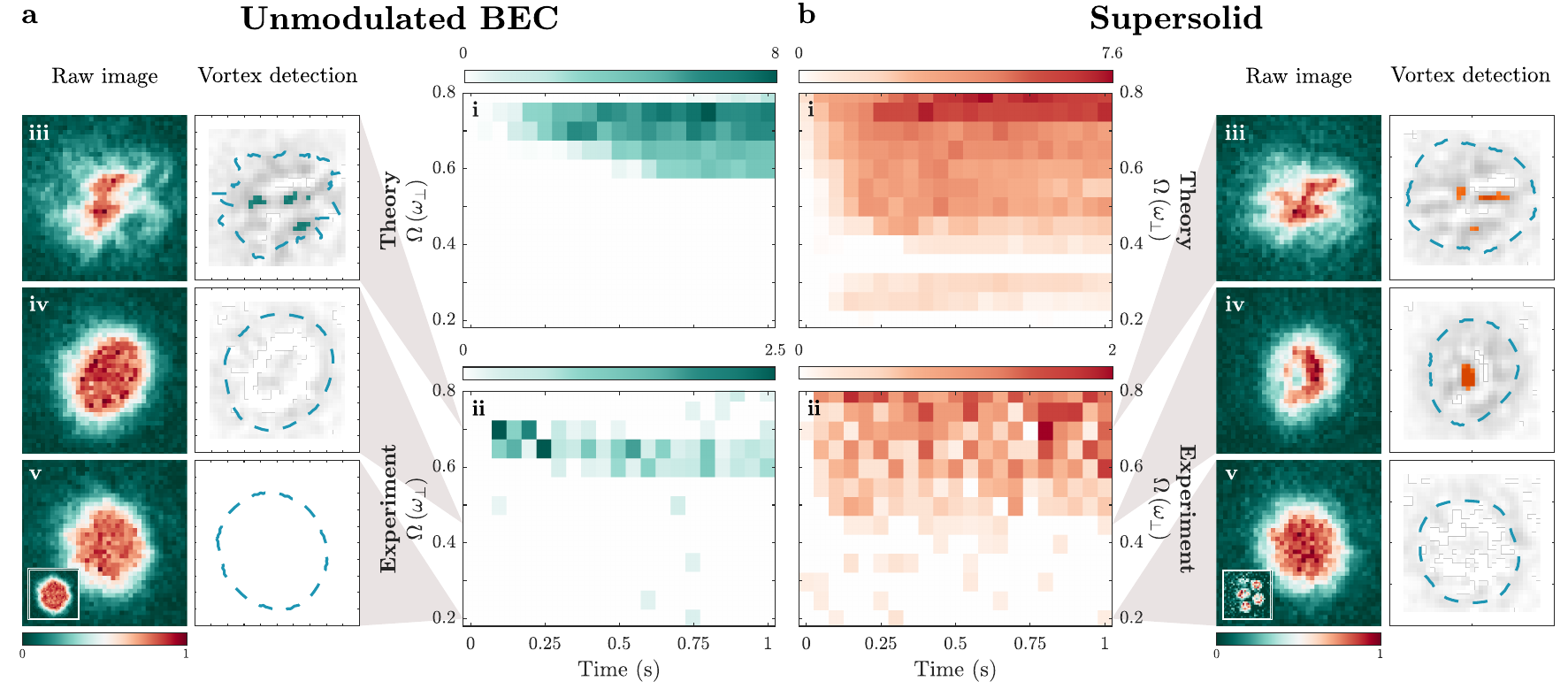}

\caption{\textbf{Vortex nucleation in a dipolar supersolid and BEC.} Average vortex number (colorbar) as a function of rotation time and $\Omega$ for an \textbf{a} unmodulated BEC and \textbf{b} supersolid. Panels show [i] the simulations, [ii] the experimental observation, where in \textbf{a} the absolute value of the magnetic field is held at $19.30(2)\,\si{G}$, but in \textbf{b} is instead ramped from $18.30(2)\,\si{G}$ to $19.30(2)\,\si{G}$ in $1\,\si{ms}$ at the end of the rotation. Exemplar images [iii-v] of normalized density taken after $250\,\si{ms}$ of rotation are shown for both cases. Detected vortices are shown in color when the residuals exceed the chosen threshold (0.34), and the condensate radius is marked by a blue dashed line, see \nameref{appendix:Methods} for more details. Insets of [v] show initial states. All images are taken after $3\,\si{ms}$ expansion, except the non-rotating supersolid state, which is a phase-contrast image with $\theta\,{=}\,0^\circ$. In the experiment, the trap has frequencies $(\omega_\perp,\,\omega_z)\,{=}\,2\pi\,{\times}\,[50.3(2),107(2)]\,\si{Hz}$, and the initial condensed atom number is $N\,{\approx}\,3\times10^4$. For the simulation: $(\omega_\perp,\,\omega_z)\,{=}\,2\pi\,{\times}\,[50,103]\,\si{Hz}$, with \textbf{a} $a_s\,{=}\,104a_0$, initial $N\,{=}\,2\,{\times}\,10^4$, and \textbf{b} $a_s\,{=}\,93a_0$, and initial $N\,{=}\,3\,{\times}\,10^4$, where three-body recombination losses have been added.}
\label{fig:Fig3} 
\end{figure*}

\noindent Based on our simulations, on the one hand, we anticipate vortex nucleation in the supersolid already at modest rotation frequencies, but on the other hand, the density modulated initial state poses a unique challenge in vortex detection. Traditional methods for probing quantized vortices in quasi-homogeneous ultracold quantum gases typically rely on observing density depletions of an expanded cloud\cite{Matthews1999via, Madison2000vfi,AboShaeer2001oov,Zwierlein2005vas}. In the context of supersolids, vortices nest within the low-density interstitial areas between the droplets, reducing the contrast\cite{Sindik2022car,Poli2023gir}. We implement an imaging sequence inspired by a recent theoretical proposal\cite{Sindik2022car} that draws parallels with a protocol employed to observe vortices in strongly interacting Fermi gases\cite{Zwierlein2005vas}. In particular, we project the SSP into the BEC phase just before releasing the atoms from the trap by rapidly ($1\,\si{ms}$) increasing the scattering length. This projection effectively “melts” the high density peaks, providing a more homogeneous density profile. Since vortices are topologically protected defects they are expected to survive during this state projection\cite{Sindik2022car}. Finally, we probe the system with vertical absorption imaging after $3\,\si{ms}$ of expansion, without allowing time for further dynamics in the BEC phase.

Figure\,\ref{fig:Fig3} summarizes our main results, where we compare the behavior of a BEC and SSP under magnetostirring. Akin to theory, we see three regimes. At low frequencies ($\Omega<\Omega^*_\mathrm{SSP}$), we do not observe vortices in either state [a(v)-b(v)]. Here, in the residuals (see \nameref{appendix:Methods}), we see the impact of the interaction quench, imparting small amplitude sound waves in the density of b(v). For $\Omega^*_\mathrm{SSP}<\Omega<\Omega^*_\mathrm{BEC}$, a striking difference between the BEC and the SSP response to rotation appears [a(iv)-b(iv)]. While the former does not show vortices, in the supersolid we clearly observe the appearance of a vortex in the central region of the cloud.
Finally, at a larger frequency ($\Omega>\Omega^*_\mathrm{BEC}$), we observe multiple vortices in both cases [a(iii)-b(iii)]. This confirms the expected reduction in vortex nucleation frequency, the first characteristic feature of the impact of supersolidity.

In what follows, we perform a systematic study as a function of $\Omega$ in order to identify the threshold nucleation values and the vortex number behavior as a function of rotation frequency. We trace the time evolution of the rotating system both in the SSP and BEC phase and extract for each time step and $\Omega$ the number of vortices. We show the average vortex number obtained for each measurement in Fig.\,\ref{fig:Fig3} (ii) together with the corresponding numerical simulations (i).
Our protocol for extracting the vortex number is detailed in the \nameref{appendix:Methods} section. We demonstrate that, while the exact vortex number depends on the specifics of the analysis, the overall qualitative behavior remains consistent.

In the unmodulated case (Fig.\,\ref{fig:Fig3}a), we observe the expected resonant behavior around $\Omega\,{\approx}\,0.7\omega_\perp$\cite{Klaus2022oov}. After $0.5\,\si{s}$ of rotation, both the experiment and theory show $\Omega^*_\text{BEC}\,{\approx}\,0.6\omega_\perp$. In the experiment, atom number losses at fast rotation frequencies can suppress the production of vortices\cite{Klaus2022oov}.
In the supersolid case (Fig.\,\ref{fig:Fig3}b), we are able to observe clear evidence for the threshold behavior for vortex nucleation. For driving frequencies greater than $\Omega\,{\approx}\,0.4\omega_\perp$, vortices persist even up to $1\,\si{s}$, and there is an increase of vortex number with rotation frequency. This behavior is in contrast to the BEC case, where above $\Omega\,{=}\,0.75\omega_\perp$ we do not observe vortices, unveiling the competing superfluid and solid contributions.

Additionally, theory  [b(v)] predicts a superfluid quadrupole resonance centered at $\Omega\,{\approx}\,0.3\omega_\perp$, with one or at most two vortices entering the cloud. A detailed analysis of the experimental data reveals a signature compatible with the existence of this resonance, see \nameref{appendix:Methods}. However, a dedicated investigation beyond the scope of this work would be required to confirm this feature.

\vspace{0.1cm}
\noindent {\bf Interference patterns}

\begin{figure}[t]
\centering
\includegraphics[width=0.9\columnwidth]{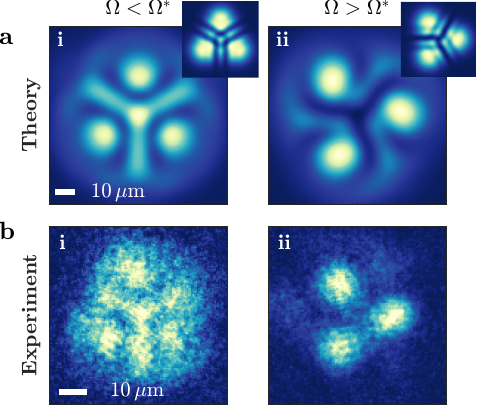}
\caption{\textbf{Time-of-flight interference pattern.} \textbf{a} $36\,\si{ms}$ real-time expansion interference pattern for three droplets (i) in the absence of a vortex and (ii) with a vortex. \textbf{b} Experimental observation after TOF (i) without rotation and (ii) after $189\,\si{ms}$ of rotation at $\Omega\,{=}\,0.3\omega_\perp$ with $\theta\,{=}\,30^\circ$, before spiraling up to $\theta\,{=}\,0^\circ$ in $11\,\si{ms}$, while $\Omega$ is kept constant. This was done to ensure that the radial droplet expansion and interference is perpendicular to the imaging axis. The supersolid is produced at $18.24(2)\,\si{G}$ with $(\omega_\perp,\,\omega_z)\,{=}\,2\pi\,{\times}\,[50.0(4),113(2)]\,\si{Hz}$, the condensed atom number $N\,{\approx}\,5\times10^4$. The theoretical parameters: $N\,{=}\,3.5\times10^4$, $(\omega_\perp,\omega_z)\,{=}\,2\pi\,{\times}\,[50,113]\,\si{Hz}$ and $a_s\,{=}\,92.5a_0$.}
\label{fig:Fig4}
\end{figure} 

\noindent The modulation of supersolid states presents a unique possibility for extracting the phase information, as the presence or absence of a vortex strongly impacts the interference pattern after time-of-flight (TOF)\cite{Gallemi2020qvi}. This is readily observable by performing expansion calculations with the eGPE, as shown in Fig.\,\ref{fig:Fig4}a. In the presence of a vortex, the interference pattern shows a pronounced minimum in the central region of the signal [a(ii)], which is clearly not the case in a vortex free supersolid [a(i)]. This remarkable feature is a direct consequence of the phase winding and can even be reproduced by a simple toy model simulating the expansion of three non-interacting Gaussian wavepackets, as shown in the insets of Fig.\,\ref{fig:Fig4}a(i) and (ii). Note that in the eGPE, the expansion time was set to $36\,\si{ms}$, during which the self-bound nature of the droplets slows down the expansion corresponding to a few $\si{ms}$ in the toy model. Furthermore, this time is strongly dependent on interaction and trap parameters, making the pattern very sensitive to parameter variations, see \nameref{appendix:Methods} for more details. Unlike vortex interference patterns from unmodulated states, there is no longer a simple hole left in the center of the cloud, but rather a three-pointed star structure reflecting the symmetries present in the density. The spiral arms appear due to the nonlinear azimuthal $2\pi$ phase winding\cite{Ancilotto2021vpi}, where between each droplet there is a line of minimum signal given by the phase difference of each droplet, in this case, $2\pi/3$. In our calculation, we opt for an initial state featuring three droplets instead of the previously used four droplet state. The symmetry of this state, characterized by equal interdroplet spacing, yields a singular and simple interference pattern when the vortex is in the center of the system, facilitating the distinction between a vortex and vortex-free state.

When performing the experiment with similar parameters as the theory, we observe a remarkable similarity across the resulting phase pattern. Figure\,\ref{fig:Fig4}b shows an example interference pattern for a non-rotating sample [b(i)] and the one for a three droplet supersolid when rotating above the theoretically obtained critical vortex nucleation frequency $\Omega^*=0.1\omega_\perp$ [b(ii)]. In the latter case, we clearly observe a signal minimum at the center, providing the observation of vortices directly in the supersolid state. To test the robustness of this observation, we repeat the measurement many times, and study the occurrence of the non-vortex [b(i)] or vortex [b(ii)] pattern. Among the images with a clear interference pattern, about 70\% contain a vortex signature when rotating above $\Omega\,{=}\,0.3\omega_\perp$, see \nameref{appendix:Methods}. The remaining fraction can be understood by considering that supersolid states exist in a very small parameter regime\cite{Poli2021msi}, and typical shot-to-shot atom number and magnetic-field ($a_s$) fluctuations can significantly alter the observed interference pattern.

\vspace{0.1cm}
\noindent {\bf Conclusions}

\noindent After three decades since the original predictions\cite{Pomeau1994doa}, we report on the observations of vortices in a supersolid state. 
This result is relevant not only because it adds the final piece to the cumulative framework of evidence for superfluidity in this state\cite{Chomaz2022dpa}, but also because it reveals a distinctive vortex behavior in the supersolid. The system's characteristic response to rotation can serve as a fingerprint to identify supersolidity in diverse systems with multiple broken symmetries, over scales ranging from solid-state systems\cite{Hamidian2016doa}, high-temperature superconductors\cite{Blatter1994vih,Kwok2016vih}, and helium platforms\cite{Nyeki2017isa,Levitin2019efa}, to a neutron star's inner crust\cite{Chamel2012nci,Poli2023gir}.

Furthermore, in the context of supersolids, a fascinating interplay of competing length scales emerges. These include the separation between vortices, the wavelength of the self-forming crystal, and the diameter of the vortex core. This competition has the potential to lead to intriguing dynamics, ranging from constrained motion and pinning to avalanche escape. These phenomena are genuinely unique to supersolids.

\vspace{0.1cm}
\noindent {\bf Acknowledgements} \\
\noindent We are indebted to Jean Dalibard for inspiring discussions on the interference pattern of supersolids in the presence of a vortex. We thank Wolfgang Ketterle, Sandro Stringari, Alessio Recati and Giacomo Lamporesi for discussions. This work was supported by the European Research Council through the Advanced Grant DyMETEr (\href{https://doi.org/10.3030/101054500}{No.\,101054500}), the QuantERA grant MAQS by the Austrian Science Fund FWF (No.\,I4391-N), a joint-project grant from the Austrian Science Fund FWF (\href{https://doi.org/10.55776/I4426}{No.\,I-4426}), and a NextGeneration EU grant AQuSIM by the Austrian Research Promotion Agency FFG (No.\,FO999896041), and by the Austrian Science Fund (FWF) Cluster of Excellence QuantA (\href{https://doi.org/10.55776/COE1}{10.55776/COE1}). A.L.~acknowledges financial support through the Disruptive Innovation - Early Career Seed Money grant by the Austrian Science Fund FWF and Austrian Academy of Science ÖAW. E.P.~acknowledges support by the Austrian Science Fund (FWF) within the DK-ALM (\href{https://doi.org/10.55776/W1259}{No.\,W1259-N27}). T.B.~acknowledges financial support through an ESQ Discovery grant by the Austrian Academy of Sciences.

\vspace{0.1cm}
\noindent {\bf Author contributions} \\
\noindent E.C., L.K., A.L., C.U., C.P., M.J.M., and F.F. performed the experimental work and data analysis. E.P. and T.B. performed the theoretical work. All authors contributed to the interpretation of the results and the preparation of the manuscript.

\vspace{0.1cm}
\noindent {\bf Data availability}\\
\noindent Data pertaining to this work can be found at \url{ https://doi.org/10.5281/zenodo.10695943}. Source data are provided with the paper.

\vspace{0.1cm}
\noindent {\bf Code availability}\\
\noindent The codes that support the findings of this study are available from the corresponding author upon reasonable request.

\vspace{0.1cm}
\noindent {\bf Competing interests}\\
\noindent The authors declare no competing interests.

\vspace{0.1cm}
\noindent {\bf Additional information}\\
\noindent {\bf Correspondence and requests for materials} should be addressed to F. Ferlaino.

\linespread{1}
%

\appendix
\label{appendix}
\counterwithin{figure}{section}
\clearpage
\noindent
\phantomsection
\vspace{-1.25cm}

\section{Methods} \label{appendix:Methods}
\renewcommand{\figurename}{Extended Data Fig.}
\renewcommand\thefigure{\arabic{figure}} 
\setcounter{figure}{0}

\noindent{\bf Experimental procedure}

\noindent We prepare an ultracold gas of $^{164}$Dy atoms in an optical dipole trap (ODT), similar to our previous work\cite{Klaus2022oov}. The trap is formed through three overlapping laser beams, operating at $1064\,\si{nm}$. All the studies are performed in a cylindrically symmetric trap, typically with $\omega_\perp\,{=}\,2\pi\,{\times}\,50.3(2)\,\si{Hz}$, where $\omega_\perp$ is the geometric average $\omega_\perp\,{=}\,\sqrt{\omega_x\omega_y}$. The aspect ratio $\omega_z/\omega_\perp$ varies from $2$ to $3$; the specific values of $\omega_z$ are stated in the figures' captions. The aspect ratio $\omega_x/\omega_y$ is crucial for the applicability of magnetostirring\cite{Klaus2022oov}: throughout the paper, the deviation of $\omega_x/\omega_y$ from $1$ is ${<}\,2\%$. 

\begin{figure}
    \centering
    \includegraphics[width=0.9\textwidth]{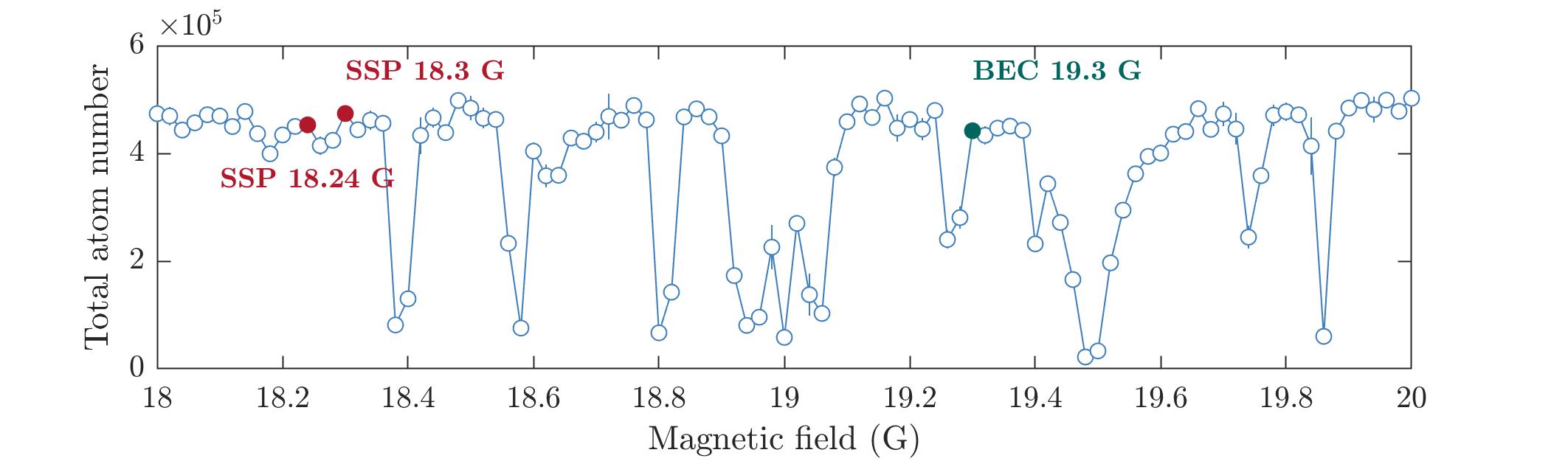}
    \caption{\textbf{Loss spectrum of $\mathbf{^{164}}$Dy}. The spectrum is obtained from horizontal absorption imaging, by varying the magnetic field at which the evaporative cooling ($T\,{\approx}\,500\,\si{nK}$) is conducted, with a step size of $20\,\si{mG}$. The magnetic field values used are highlighted in red (SSP) and green (BEC). Error bars represent the standard error.}
    \label{fig:extendedFig1}
\end{figure}

For this work, we tilt the magnetic field vector $\mathbf{B}$ from the vertical position to $\theta\,{=}\,30^\circ$ from the $z$-axis in the last stage of evaporation, while maintaining its magnitude constant. The values of the magnetic field are: $19.30(2)\,\si{G}$ for the unmodulated BEC, $18.30(2)\,\si{G}$ for the SSP in Fig.\,\ref{fig:Fig3}, and $18.24(2)\,\si{G}$ for Figs.\,\ref{fig:Fig2} and \ref{fig:Fig4}. The small change in magnetic field is required to maintain supersolidity between the two respective choices of tight trapping frequency. The magnetic field is calibrated through radio frequency (RF) spectroscopy. Moreover, $^{164}$Dy has a dense spectrum of narrow Feshbach resonances, as shown in Extended Data Fig.\,\ref{fig:extendedFig1}. We use the positions of such resonances as references to compensate for drifts of the magnetic field. The condensed atom number after the evaporation sequence ranges from $3\,{\times}\,10^4$ to $7\,{\times}\,10^4$, depending on the measurement. 

After preparation, the magnetic field is rotated; details can be found in the following sections. Finally, we image the quantum gas using a $421\,\si{nm}$ light pulse, propagating along the $z$-axis. For the data in Figs.\,\ref{fig:Fig2} and \ref{fig:Fig3}, we let the atomic cloud expand for $3\,\si{ms}$ and take a phase contrast and absorption image, respectively. When comparing theoretical and experimental images, we rescale the image size by $1.36$ in the theory to account for this small expansion time. The results of Fig.\,\ref{fig:Fig4} are instead obtained with absorption imaging after $36\,\si{ms}$ TOF, and following an $11\,\si{ms}$ spiral up to $\theta\,{=}\,0^\circ$. This was done to ensure that the radial droplet expansion and interference is perpendicular to the imaging axis.

For the experimental images in Fig.\,\ref{fig:Fig2}, we enhanced the contrast of the droplets by applying a Gaussian filter of size $\sigma\,{=}\,1$ px (${\simeq}\, 0.5\,\mu\si{m}$) followed by a sharpening convolution filter with kernel $F$:
\begin{equation}    
F=\begin{bmatrix}
0 & -1 & 0\\
-1 & 5 & -1\\
0 & -1 & 0
\end{bmatrix}.\label{eqn:filter}\end{equation}

\vspace{0.1cm}
\noindent{\bf Magnetostirring}

\noindent To magnetostir the atomic cloud, we rotate the magnetic field vector around the $z$-axis\klaus. In brief, the breaking of cylindrical symmetry that enables the transfer of angular momentum by rotating the magnetic field vector $\mathbf{B}$ (magnetostirring) is achieved by tilting $\mathbf{B}$ into the plane. This is a direct consequence of the phenomenon of magnetostriction\cite{Stuhler2005ood}. For all the measurements in this paper, $\mathbf{B}$ is tilted from the $z$-axis by an angle $\theta\,{=}\,30^\circ$. At our magnetic field values, this angle is optimal for vortex nucleation within the experimental time scales\cite{Bland2023vid}. In general, smaller angles would increase the nucleation time; at the same time, a much bigger angle would make the dipolar interaction dominantly attractive, holding the cloud together and thus also increasing the nucleation time. From the experimental point of view, $\theta\,{=}\,30^\circ$ enables the observation of the droplets aligning along $\mathbf{B}$ while retaining the ability to discern individual droplets when observing the integrated density, see Fig.\,\ref{fig:Fig2}. For all datasets, we then directly rotate $\mathbf{B}$ at the chosen frequency $\Omega$. The rotation is continued for a rotation time $t_\Omega$ after which the ODT is turned off, and an image is taken after expansion.

\vspace{0.1cm}
\noindent{\bf Scattering length}

\noindent The conversion from magnetic field to scattering length for $^{164}$Dy at our magnetic field values has not been mapped. However, combining knowledge on the conversion in other magnetic field ranges\cite{Tang2015sws, Maier2015buf, Tang2016aeo}, together with the theoretical identification of the critical scattering lengths for the BEC to SSP transition, allows for an educated guess. It is important to highlight that the isotope $^{164}$Dy has the advantage of exhibiting supersolidity at the background value of the scattering length, while the BEC phase usually requires some mild tuning of $a_s$. The specific values used in this paper are highlighted on the Feshbach loss spectrum in Extended Data Fig.\,\ref{fig:extendedFig1}. For our theoretical simulations (see below), we find that a scattering length $a_s$ in the range $90a_0$-$95a_0$ gives a good agreement with the experimentally observed supersolid states.

\vspace{0.1cm}
\noindent{\bf Interaction quench}

\noindent For the in situ detection of vortices in the supersolid phase, we map the supersolid into an unmodulated BEC, similar to the approach used to observe them in the BCS phase of strongly interacting Fermi gases\cite{Zwierlein2005vas}. In particular, we increase the absolute value of the magnetic field from $18.30(2)\,\si{G}$ to $19.30(2)\,\si{G}$ in $1\,\si{ms}$ after stopping the rotation, and we then release the sample from the trap for $3\,\si{ms}$ before imaging. We repeat this sequence for different values of angular velocity $\Omega$ and for different rotation times $t_\Omega$. For each experimental point in Fig.\,\ref{fig:Fig3}a and \ref{fig:Fig3}b, we take 7-9 pictures. Using phase contrast imaging, we ensured that the ramp time is long enough to melt the droplets into an unmodulated state, but also short enough to avoid atom losses when crossing the Feshbach resonances present between the initial and final magnetic field values (see Extended Data Fig.\,\ref{fig:extendedFig1}).

\vspace{0.1cm}
\noindent{\bf Extended Gross-Pitaevskii equation}

\noindent At the mean-field level, the ground state solutions, time-dependent dynamics, and nature of the BEC-to-SSP transitions are well described by the extended Gross-Pitaevskii formalism\cite{Waechtler2016qfi,FerrierBarbut2016ooq, Chomaz2016qfd,Bisset2016gsp}. This combines the two-body particle interactions, described by the two-body pseudo-potential, 
\begin{eqnarray}
    U(\textbf{r}) = \frac{4\pi\hbar^2a_{\mathrm s}}{m}\delta(\textbf{r})+\frac{3\hbar^2a_\text{dd}}{m}\frac{1-3\left(\hat{\mathbf{e}}(t)\cdot\hat{\mathbf{r}}\right)^2}{r^3}\,,
\end{eqnarray}
where the first term describes short-range interactions governed by the s-wave scattering length $a_s$, with Planck's constant $\hbar$ and particle mass $m$. This quantity is independently tunable through Feshbach resonances. The second term represents the anisotropic and long-ranged dipole-dipole interactions, characterized by dipole length $a_\text{dd}\,{=}\,\mu_0\mu_m^2m/12\pi\hbar^2$, with magnetic moment $\mu_m$ and vacuum permeability $\mu_0$. We always consider $^{164}$Dy, such that $a_\text{dd}\,{=}\,130.8\,a_0$, where $a_0$ is the Bohr radius. For the trap parameters and atom numbers used here, the supersolid phase is found for scattering lengths in the range $a_s\,{=}\,[90,95]a_0$, i.e.\,$\epsilon_\text{dd}\,{=}\, a_{\mathrm{dd}}/a_{\mathrm s}\ge1.37$. The dipoles are polarized uniformly along a time-dependent axis, given by
\begin{eqnarray}
    \hat{\mathbf{e}}(t)=(\sin\theta(t)\cos\phi(t),\sin\theta(t)\sin\phi(t),\cos\theta(t))\,
\end{eqnarray}
with time dependent polarization angle $\theta(t)$ and $\phi(t)\,{=}\,\int_0^t \text{d}t'\Omega(t')$, for rotation frequency protocol $\Omega(t)$.

Three-body recombination losses are prevalent in dipolar supersolid experiments due to the increased peak density when compared to unmodulated states. In the theory, these are introduced through a time-dependent atom loss
\begin{eqnarray}
    \dot{N} = -L_3\langle n^2 \rangle N\,,
\end{eqnarray}
for density $n$. We take the fixed coefficient $L_3\,{=}\,1.2\,{\times}\, 10^{-41}$m$^{6}$s${^{-1}}$ for our simulations\cite{FerrierBarbut2016ooq}. This leads to an additional non-Hermitian term in the Hamiltonian $-\text{i}\hbar L_3n^2/2$.

Beyond-mean-field effects are treated through the inclusion of a Lee–Huang–Yang correction term\cite{Lima2011qfi}
\begin{eqnarray}
    \gamma_\text{QF}=\frac{128\hbar^2}{3m}\sqrt{\pi a_s^5}\,\text{Re}\left\{ \mathcal{Q}_5(\edd) \right\} \, ,
\end{eqnarray}
where $\mathcal{Q}_n(x)=\int_0^1 \text{d}u\,(1-x+3xu^2)^{n/2}$, which has an imaginary component for $x>1$. Finally, the full extended Gross-Pitaevskii equation (eGPE) then reads\cite{Waechtler2016qfi,FerrierBarbut2016ooq, Chomaz2016qfd,Bisset2016gsp}
\begin{eqnarray}
    \text{i}\hbar\frac{\partial\psi(\textbf{r},t)}{\partial t} =  \bigg[-\frac{\hbar^2\nabla^2}{2m} +V_\text{trap}-\frac{\text{i}\hbar L_3}{2}|\psi(\textbf{r},t)|^4   \nonumber\\
    +\int\text{d}^3\textbf{r}'\, U(\textbf{r}-\textbf{r}')|\psi(\textbf{r}',t)|^2  +\gamma_\text{QF}|\psi(\textbf{r},t)|^3\bigg]\psi(\textbf{r},t)\,,
    \label{eqn:GPE}
\end{eqnarray}
where $\omega_{x,y,z}$ are the harmonic trap frequencies in $V_\text{trap}\,{=}\,\frac12m\left(\omega_x^2x^2+\omega_y^2y^2+\omega_z^2z^2\right)$. The wavefunction $\psi$ is normalized to the total atom number $N\,{=}\,\int {\mathrm d}^3\mathbf{r}|\psi|^2$. Stationary solutions to Eq.\,\eqref{eqn:GPE} are found through the standard imaginary time procedure. The initial state $\psi(\textbf{r},0)$ of the real-time simulations is obtained by adding non-interacting noise to the stationary solution $\psi_0(\textbf{r})$. Given the single-particle eigenstates $\phi_n$ and the complex Gaussian random variables $\alpha_n$ sampled with $\langle|\alpha_n|^2\rangle\,{=}\, (e^{\epsilon_n/k_BT}-1)^{-1}+\frac12$ for a temperature $T\,{=}\,20\,\si{nK}$, the initial state can be described as $\psi(\textbf{r},0)\,{=}\, \psi_0(\textbf{r}) + \sum_n' \alpha_n \phi_n(\textbf{r})$,  where the sum is restricted only to the modes with $\epsilon_n\,{\le}\,2k_BT$\cite{blakie2008das}.

\vspace{0.1cm}
\noindent{\bf Choosing simulation parameters}

\noindent In the theory there are two parameters known to high precision, the atomic mass and the dipolar strength, and seven parameters that are only known within broad error bars, $\{a_s, L_3,\omega_{x,y,z},T,N\}$. Here, we explore the impact of varying these parameters in the theory on the main results of the paper, and justifying their use when comparing to the experimental data.

\begin{figure}[t]
\centering    
\includegraphics[width=0.8\textwidth]{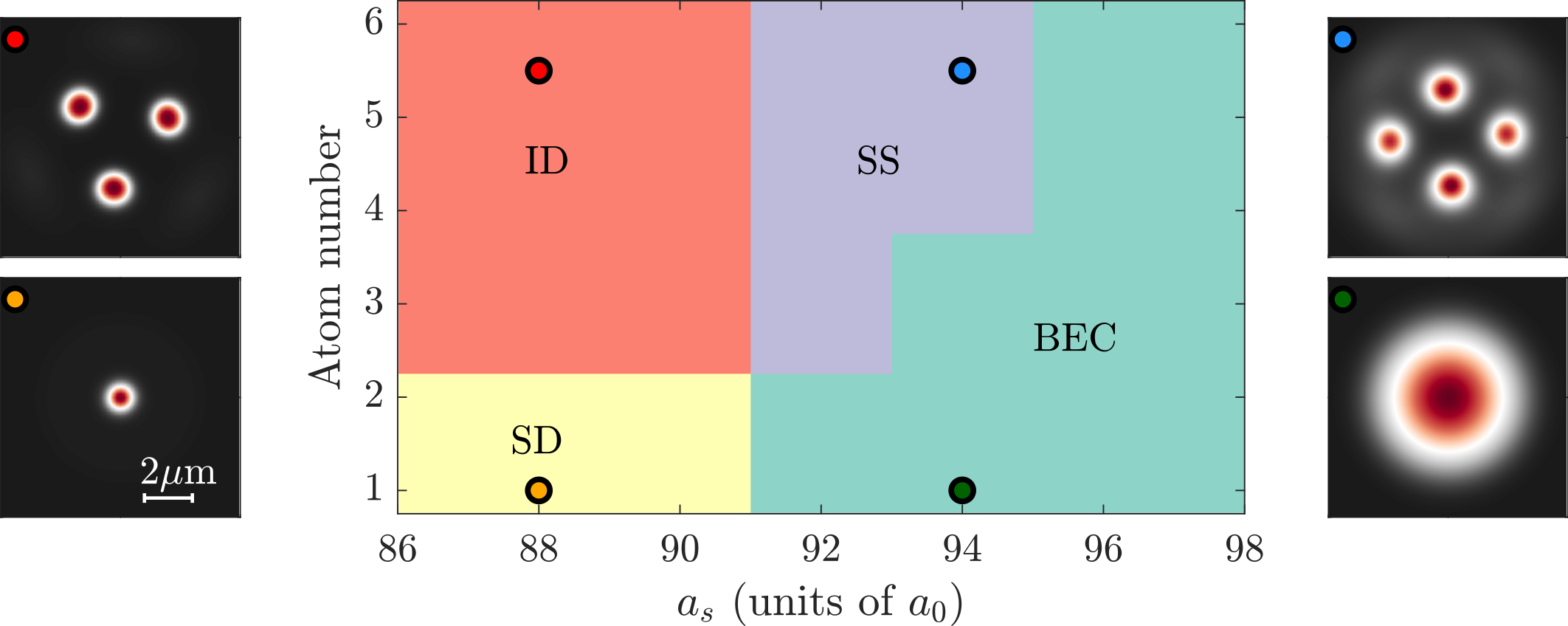}
    \caption{\textbf{Ground state phase diagram obtained varying the atom number and the scattering length}. The results are obtained from eGPE calculations with $(\omega_\perp,\,\omega_z)\,{=}\,2\pi\,{\times}\,[50,103]\,\si{Hz}$. The identified phases are: BEC, SS (supersolid), SD (single droplet) and ID (isolated droplets). On the sides, exemplar ground states extracted from the phase diagram.}
    \label{fig:extendedFig_phasediagram}
\end{figure}

In Extended Data Fig.\,\ref{fig:extendedFig_phasediagram} we show the ground state phase diagram for different values of atom number and scattering length $a_s$. We identify four different phases: unmodulated BEC, SSP (supersolid), ID (isolated droplets) and SD (single droplet). To characterize the different phases we consider the density contrast $C=(n_{max}-n_{min})/(n_{max}+n_{min})$, where $n$ is the column density. In the SSP and ID regime, the contrast is calculated as the average value between each pair of droplets. States with more than one droplet and with a contrast larger than the threshold value $0.98$ are identified as ID. Rotation of the ID state is discussed in detail in the next section.

\begin{figure}[t]
\centering    
\includegraphics[width=\textwidth]{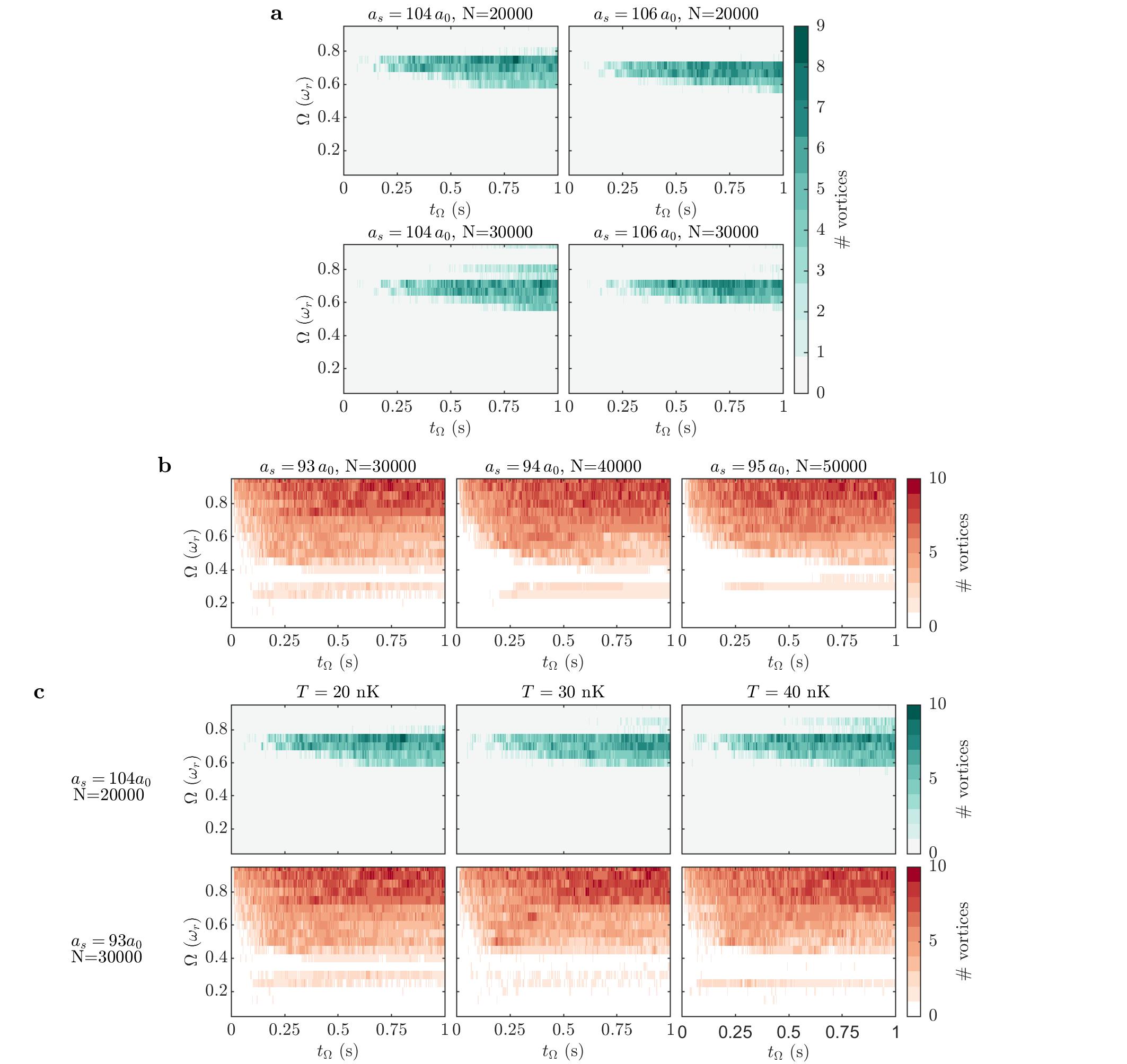}
    \caption{\textbf{Vortex nucleation in a dipolar BEC and supersolid for different parameters}. \textbf{a} Vortex nucleation in a dipolar BEC and \textbf{b} in a supersolid, for different atom number and different scattering length $a_s$. \textbf{c} Vortex nucleation for initial noise with different temperatures. All the results are obtained from eGPE calculations with $(\omega_\perp,\,\omega_z)\,{=}\,2\pi\,{\times}\,[50,103]\,\si{Hz}$, magnetic-field angle from the $z$-axis $\theta\,{=}\,30^\circ$ and 3-body recombination losses are included.}
    \label{fig:extendedFig_3}
\end{figure}

In Extended Data Fig.\,\ref{fig:extendedFig_3}a, we show how the rotational response of a BEC is robust against atom number and scattering length variations. Varying these parameters slightly moves the exact position and size of the resonance window for vortex nucleation, and although the precise values are difficult to pin down, all of our simulations show resonance behavior. 

In Extended Data Fig.\,\ref{fig:extendedFig_3}b, we show how the rotational behavior of a supersolid is robust against atom number variations. In this case, the scattering length $a_s$ was also tuned to maintain supersolidity. As in the BEC case, varying these parameters slightly moves the exact position of the resonance and of the threshold but the overall structure of the rotational response remains consistent.

We finally consider the rotational response of both BEC and SSP for different temperatures. In Extended Data Fig.\,\ref{fig:extendedFig_3}c we compare $T = 20$nK, which is the case in the main text, to $30$nK and $40$nK. In the BEC case, the key difference is that the higher temperature broadens slightly the resonance window. In the supersolid case, the higher temperature obscures the small resonance at smaller $\Omega$. Both of these observations are in keeping with the experimental results, and provide new perspectives on the observed differences between theory and experiment, where the resonance at smaller $\Omega$ may be better observable at smaller temperatures. Crucially, the physics is independent of the choice of the initial noise.

\vspace{0.1cm}
\noindent{\bf Isolated droplet regime}

\noindent For values of the scattering length $a_s$ lower than the ones required to have a supersolid state, the superfluid connection between crystal sites disappears and the system enters the so-called isolated (or independent) droplet (ID) regime. In this regime, the system does not exhibit global phase coherence and each droplet evolves as an independent BEC.

The ID regime is not accessible in our experiment for long trapping times. In the experiment, we have chosen $^{164}$Dy, which, to our best knowledge, is the only isotope showing long-lived supersolid states with a lifetime up to \si{1}{s} \cite{Sohmen2021bla}. The long lifetime is essential for prolonged rotation and vortex seeding. This particular isotope shows long-lived supersolidity due the fortunate coincidence that the background scattering length is the one required for supersolidity in our trap, without any need of Feshbach tuning \cite{Chomaz2019lla}. Instead, to produce a BEC we need to move between two overlapping Feshbach resonances giving the right modulated background value, where here a moderate tuning is needed and still three-body losses are modest. To create an ID state, we need instead to go closer to a resonance and three-body losses destroy the state in a timescale of $100\,\si{ms}$, too short for applying the rotation protocol.  

\begin{figure}[t]
\centering    
\includegraphics[width=0.9\textwidth]{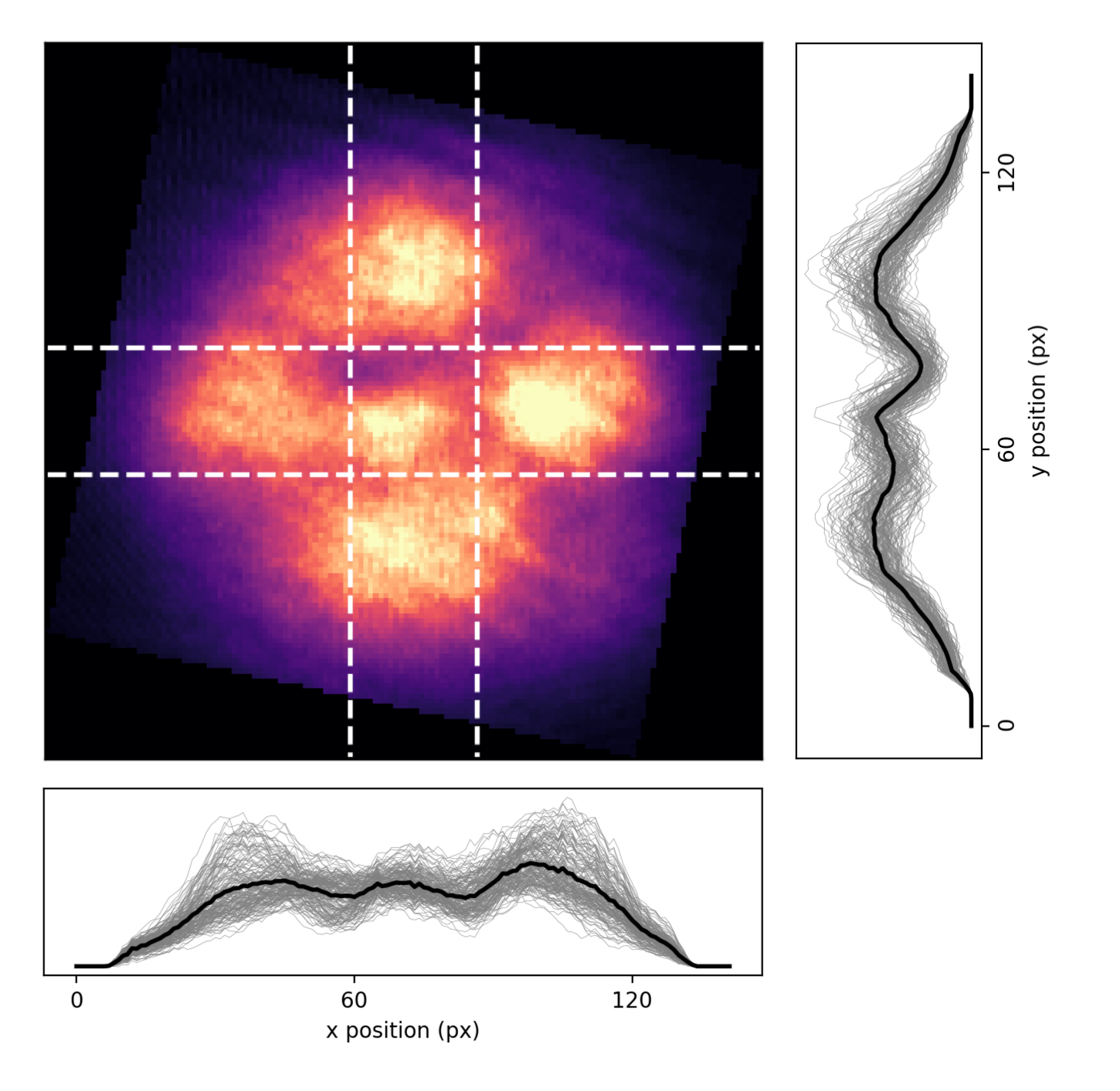}
    \caption{\textbf{Phase coherence measurement of the initial four droplet state before rotation, after $\mathbf{36\,\si{ms}}$ TOF.} The lower (right) figure shows the horizontal (vertical) integrated density. The modulation and central interference peak are present on single images (grey lines) and remain after averaging over 173 images (black line).}
    \label{fig:extendedFig_coherence}
\end{figure}

In Extended Data Fig.~\ref{fig:extendedFig_coherence} we show the phase coherence measurements that prove the initial state used for the dynamics in Fig.\,\ref{fig:Fig3} is in the supersolid regime. In Ref.\,\cite{Gallemi2020qvi} the supersolid robustness to rotation has been theoretically demonstrated. At relatively small rotation frequencies, $\Omega\lesssim0.2\omega_\perp$, the angular momentum dependence was shown to be linear in the absence of vortices, where vortices present as jumps in the total angular momentum. The linear regime is only possible if the moment of inertia of the supersolid, i.e. the density structure and superfluid connection, remains unchanged in response to rotation. Moreover, at higher rotation frequencies, rotation acts as centrifugal effect that lowers the peak density of the state, thus favoring the superfluid density connections between the droplets. The only system parameter that can change during the rotation protocol is the atom number due to losses, and lowering this will bring the state towards the BEC state, as shown in Extended Data Fig.\,\ref{fig:extendedFig_phasediagram}. However, as shown in Fig.\,\ref{fig:Fig2}, the supersolid survives for a much longer time than the one needed for seeding vortices.

In the ID regime the droplets are fully separated and there is no density between them, so the concept of vortices cannot be applied. Moreover, vortices also not able to enter the droplets themselves, as the density is very high and the self-bound nature suppresses the vortex nucleation \cite{Bland2023vid}, and even if imprinted in a droplet, they are known to be unstable \cite{Cidrim2018vis,Lee2018eoa,Lee2021nco,Li2024sav}. 

It is worth to discuss whether it is possible that in a vortex free rotating ID state, a signature of vortices emerge due to the melting process, caused by the interference of droplets with different phases. In principle this mechanism is possible, but it will occur with small probability since the random phase scrambling must match 2$\pi$ \cite{Scherer2007vfb}. Moreover, this case is completely independent of the rotation protocol and frequency, it would even happen without rotating at all. Considering our protocol and geometry, with a tilt of $30^\circ$ and melting dynamics of 4\,\si{ms}, the hypothetical vortex line would be unstable because it will be diagonal to the tight trap confinement, and will be integrated out during our absorption imaging protocol.

Another perspective pertains to the conservation of angular momentum during the quench protocol. As we keep the magnetic field both tilted and static during the interaction quench and time-of-flight, the system remains asymmetric, and there is no conservation of angular momentum. Furthermore, an azimuthally asymmetric stirred superfluid exhibits non-zero angular momentum without the need for vortices, in order to fulfill the irrotational velocity condition\cite{Stringari2016bec}.

\begin{figure}[t]
\centering    
\includegraphics[width=1\textwidth]{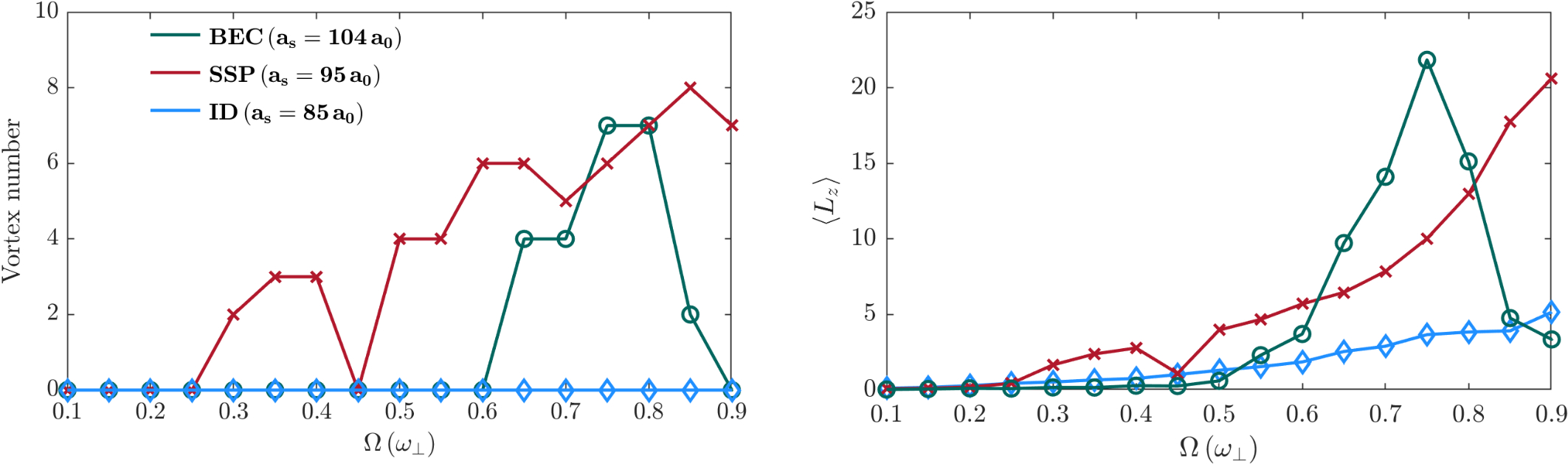}
    \caption{\textbf{Vortex number and expectation value of the angular momentum.} Left: vortex number after 1\,s of rotation. Right: expectation value of the angular momentum operator also after 1\,s of rotation. The other parameters are the same as Fig.\,\ref{fig:Fig1} of the main text.}
    \label{fig:extendedFig_Lzdiffas}
\end{figure}

In Extended Data Fig.\,\ref{fig:extendedFig_Lzdiffas} we plot the expectation value of the angular momentum operator measured for the wave function after $1\,\si{s}$ of rotation. 
In the ID regime ($a_s=85\,a_0$) the angular momentum monotonically increases as a function of $\Omega$. On the contrary, in both the BEC ($a_s=104\,a_0$) and SS ($a_s=95\,a_0$) regimes there are maxima corresponding to vortex nucleation, or, as is the case for the SS regime, a sum of both the solid-like contribution and vortices. Notice that when the vortex number is zero at $\Omega = 0.45\omega_r$ the angular momenta are equal for both the ID and supersolid phases, due to the similar density distributions. 

Finally, it is worth discussing what is the behavior of isolated droplets during long time-of-flight measurements, following the same protocol of Fig.\,\ref{fig:Fig4}. First of all, the phase of each droplet is initially random and the rotation would maintain their random character (the effect of a phase gradient across each droplet due to rotation is negligible). As a consequence of this, the produced interference pattern will not be repeatable over many experimental shots and it would be independent on the rotation frequency, in contrast to the observation of Extended Data Fig.~\ref{fig:extendedFigTOF}.
Furthermore, the expansion time required for the droplets to unbind in this geometry is longer than \SI{36}{ms} because of their stronger self-bound character. This time-of-flight expansion would produce a magnified version of the isolated droplet ground state without any low density halo around or interference structure in the middle.\\

\vspace{0.1cm}
\noindent{\bf Toy model interference pattern}

\begin{figure}[t]
\centering    
\includegraphics[width=0.9\textwidth]{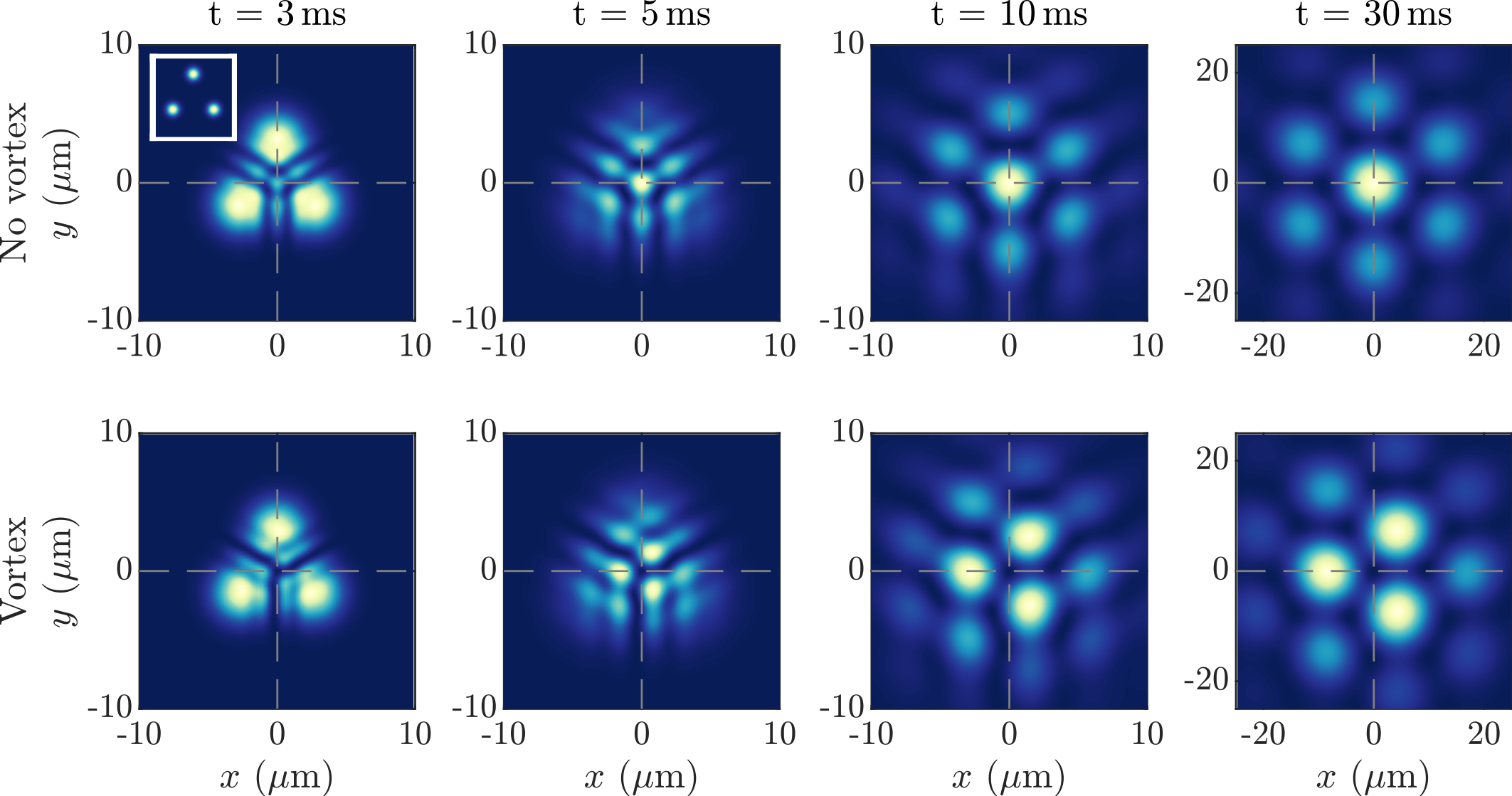}
    \caption{\textbf{Time of flight predictions from the Gaussian toy model}. Longer TOF density profiles for the solution shown in Fig.\,\ref{fig:Fig4} of the main text. The inset of the first figure shows the initial condition for all states. After $10\,\si{ms}$ the density pattern has frozen into the momentum distribution of the initial cloud. The gray lines show the axis center (0,0), highlighting the immediate difference between a no vortex and vortex expansion from the central density.}
    \label{fig:extendedFig2}
\end{figure}

\noindent Taking $N_D$ static Gaussian wavepackets with parameters of the $j^\text{th}$ wavepacket given by the widths $\bm{\sigma}_j\,{=}\,(\sigma_{1,j},\sigma_{2,j},\sigma_{3,j})$, positions $\textbf{r}_j^0\,{=}\,(r^0_{1,j}, r^0_{2,j}, r^0_{3,j})$, atom numbers $N_j$, and phase $\phi_j$, the initial total wavefunction is 
\begin{eqnarray}\label{eqn:wavpackt0}
    \psi(\textbf{r},0) &=& \sum_j^{N_D}\sqrt{\frac{N_j}{\left(2\pi\right)^{3/2}}}\exp\left({\text{i}\phi_j}\right) \\
    &\times&\prod_{k={1,2,3}}\sqrt{\frac{1}{\sigma_{k,j}}}\exp\left[-\frac14\left(r_k-r^0_{k,j}\right)^2/\sigma_{k,j}^2\right]\,.\nonumber
\end{eqnarray}
On the assumption that these wavepackets are non-interacting, then their expansion due to kinetic energy alone can be analytically calculated by applying the free particle propagator in three dimensions, such that the time-dependent solution is
\begin{eqnarray}
    \psi(\textbf{r},t) = \int_{-\infty}^{\infty}\text{d}^3\textbf{r}'\,\psi(\textbf{r}',0)K(\textbf{r},t;\textbf{r}',0)\,,
    \label{eqn:prop}
\end{eqnarray}
where
\begin{eqnarray}
    K(\textbf{r},t;\textbf{r}',t_0)=\left(\frac{m}{2\pi\text{i}\hbar(t-t_0)}\right)^{3/2}\exp\left(\frac{\text{i}m(\textbf{r}-\textbf{r}')^2}{2\hbar(t-t_0)}\right)\,.
\end{eqnarray}
Applying Eq.\,\eqref{eqn:prop} to Eq.\,\eqref{eqn:wavpackt0} gives the time-dependent multi-wavepacket solution. For brevity it is not stated here, but the exact solution transpires to be a simple time-dependent replacement of the widths $\left\{\sigma_{k,j}\,{\to}\,\sigma_{k,j}\sqrt{1+\text{i}\hbar t/(2m\sigma_{k,j}^2)}\right\}$ appearing in Eq.\,\eqref{eqn:wavpackt0}. An example of the evolution of the TOF pattern is shown in Extended Data Fig.\,\ref{fig:extendedFig2} with the parameters of Fig.\,\ref{fig:Fig4} for longer times. Note that the 3ms TOF pattern, equivalent to the $36\,\si{ms}$ when simulating the eGPE (i.e.~including interactions), has not yet evolved into the momentum distribution.

\vspace{0.1cm}
\noindent{\bf Quadrupole modes calculation}

\noindent We employ real-time simulations with the extended Gross-Pitaevskii equation to investigate the quadrupole mode frequency of the system with the tilted magnetic field, both in the BEC and in the supersolid phase. We initially perturb the system with a sudden small quadrupolar deformation of the trap and, then, we let the system evolve for $1\,\si{s}$. The deformation is done by increasing (decreasing) the trap frequency by $0.5\,\si{Hz}$ in the $x$-direction ($y$-direction) for $1\,\si{ms}$ and then restoring the trap to the original value. During the time evolution, the density distribution in the slice $z\,{=}\,0$ is fitted with a Gaussian profile, from which we extract the time-dependent width of the system during the evolution. The Fourier transform of the time-dependent width gives the frequency spectrum of all the expected superfluid and crystal quadrupole modes excited by the sudden deformation\cite{Gallemi2020qvi,Norcia2022cao}. These frequencies are in agreement with the features of the rotational response of the BEC and supersolid discussed in the main text. 

\vspace{0.1cm}

\noindent{\bf In-situ vortex detection algorithm}

\noindent To count the number of vortices, we identify the number of voids in the density in the in-situ images, following a similar procedure of our earlier work\cite{Klaus2022oov}, the steps of which are shown in Extended Data Fig.\,\ref{fig:extendedFigResiduals}. In short, we first apply a Gaussian filter of size $\sigma\,{=}\,1$ px (${\simeq}\, 0.5\,\mu\si{m}$), then the sharpening convolution filter of Eq.\,\eqref{eqn:filter} to each image $n_\mathrm{img}$ for noise reduction. We then prepare a blurred reference image $n_\mathrm{ref}$ by applying a Gaussian filter of size $\sigma\,{=}\,3\,$px~(${\simeq}\, 1.5\,\mu\si{m}$) to each $n_\mathrm{img}$ and calculate the residuals between this reference and the original image $n_\mathrm{res}\,{=}\,n_\mathrm{ref}-n_\mathrm{img}$. Finally, vortices are detected as peaks in the residual image $n_\mathrm{res}$ using a peak detection algorithm (peak\_local\_max from the \textsc{skimage} Python library). To avoid spurious vortex detection, we discard peaks with a distance below $3$ px, and peaks with an amplitude below a chosen contrast threshold of $0.34$.

We verify the robustness of the vortex detection by varying this contrast threshold between $0.34$ and $0.42$, which changes the number of selected peaks but gives the same qualitative result on the whole data set (see Extended Data Fig.\,\ref{fig:extendedFigIntegrated}). We remark that any choice of threshold will not completely remove all false positive detections, but this method allows for an unbiased measure impartially applied to all images. In the experimental data (Extended Data Fig.\,\ref{fig:extendedFigIntegrated}) there is a small peak centered at $\Omega\,{=}\, 0.35\omega_\perp$ for all thresholds considered, hinting towards the expected superfluid quadrupole mode resonance, see Fig.~\ref{fig:Fig3}.
\vspace{0.1cm}

\noindent{\bf Comparison between the vortex number in theory and experiment}

\noindent The vortex nucleation study conducted in Fig.\,\ref{fig:Fig3}, shows a mismatch in the detected vortex number between the experiment and our simulations, despite the general agreement between theory and experiment on the dependence on rotation frequency and rotation time. This disagreement is caused by the different detection methods and by experimental fluctuations, as explained in the following.

In Figs.\,\ref{fig:Fig1} and \ref{fig:Fig3}, the number of vortices is determined by counting 2$\pi$ windings in the central slice of the phase, $\text{arg}(\psi(x,y,z\,{=}\,0))$. We restrict the search to a circle of radius $6\,\mu\si{m}$, such that vortices are only counted inside the condensate surface in the BEC case, or within the halo in the supersolid state. To visualize the vortex tubes plotted in Fig.\,\ref{fig:Fig1}, we plot isosurfaces of the velocity field.

\begin{figure}[t]
    \centering
    \includegraphics[width=0.9\textwidth]{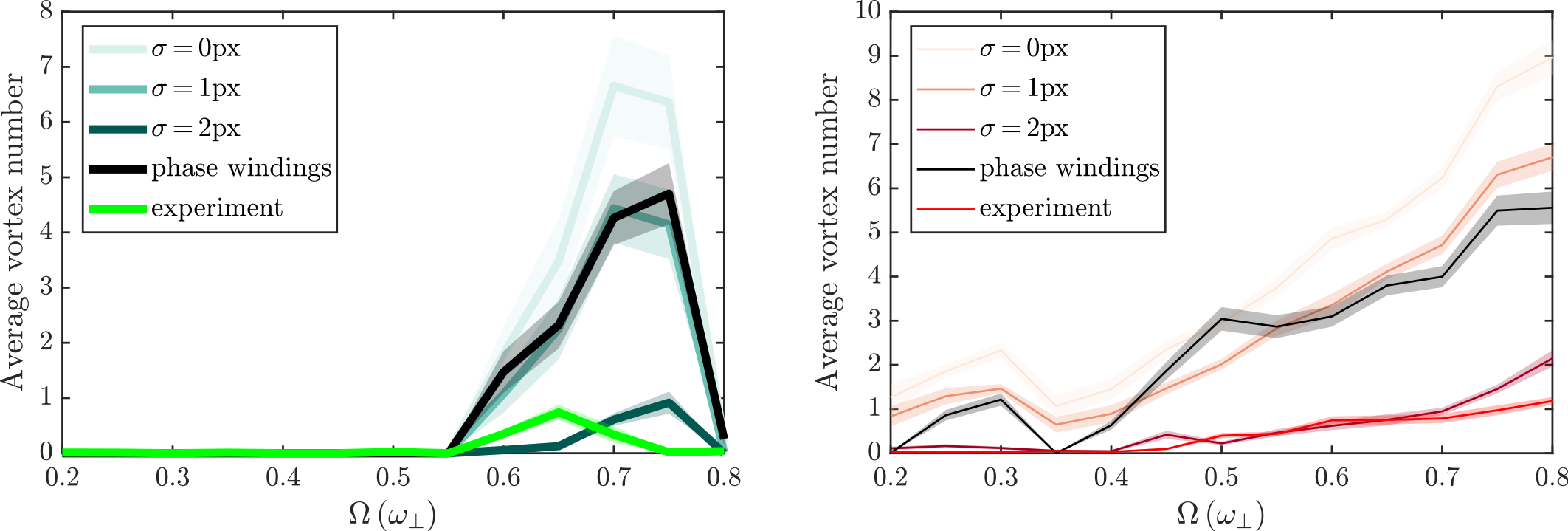}
    \caption{\textbf{Comparison of different vortex detection methods applied to the theoretical data.} 
    Each point is obtained by applying the experimental vortex detection algorithm to the states of Fig.\ref{fig:Fig3} and averaging over time. For the SSP the scattering length is ramped from $a_s=93a_0$ to $a_s=104a_0$ in $1\,\si{ms}$ and the state is expanded for $3\,\si{ms}$, before applying the algorithm. The results are shown for different size of Gaussian filter $\sigma$ and compared to the standard method of counting the $2\pi$ phase windings (black line) and the experimental data, in green (red) for the BEC (SSP). The shaded area indicates the error on the mean.}
    \label{fig:extendedFig_melted_phase_diagram}
\end{figure}

\begin{figure}[t]
\centering    
\includegraphics[width=0.9\textwidth]{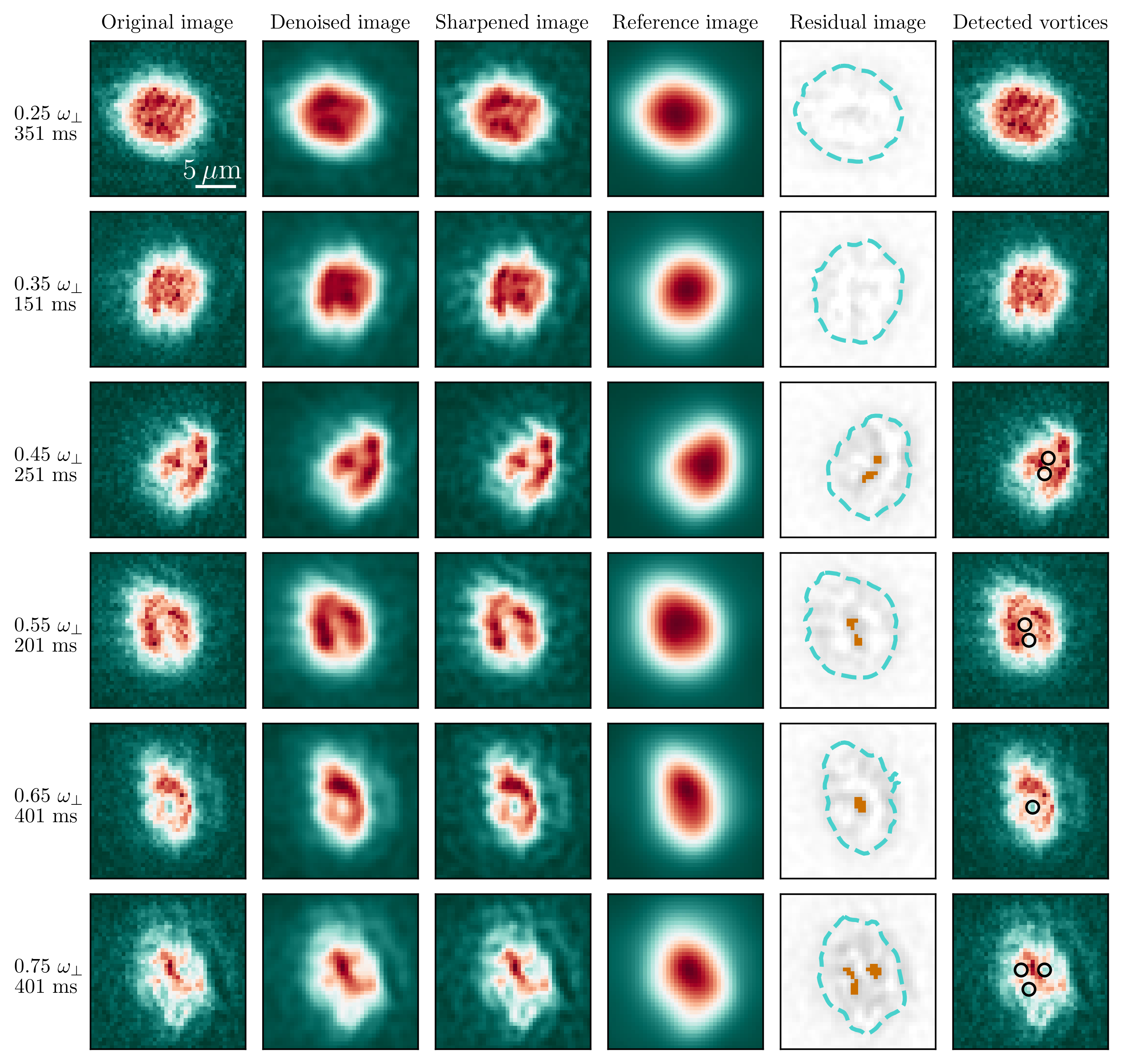}
    \caption{\textbf{Image processing for the detection of vortices.} Each row indicates different rotation frequency and duration parameters (indicated on the left), where images are taken following an interaction quench from the supersolid to unmodulated BEC phase. Each column is a step of the processing protocol which goes as follows. The data (column 1) is normalized and denoised with a Gaussian filter of size $\sigma\,{=}\,1$ (column 2), and a sharpening mask is applied to magnify the presence of vortices (column 3). The reference image is built from the data image where all density variations are eliminated with a Gaussian filter of size $\sigma\,{=}\,3$ (column 4). The residuals (column 5)  are obtained from the subtraction of the data to the reference, converting the density depletions to a positive signal. The vortices (black circles) are detected with a peak detection algorithm with threshold $0.38$. The last column shows the location of the vortices on the original image data. Varying the threshold value modifies the absolute vortex count of each individual image but not the overall qualitative result (see Extended Data Fig.\,\ref{fig:extendedFigIntegrated}).}
    \label{fig:extendedFigResiduals}
\end{figure}

\begin{figure}[t]
\centering    
\includegraphics[width=\textwidth]{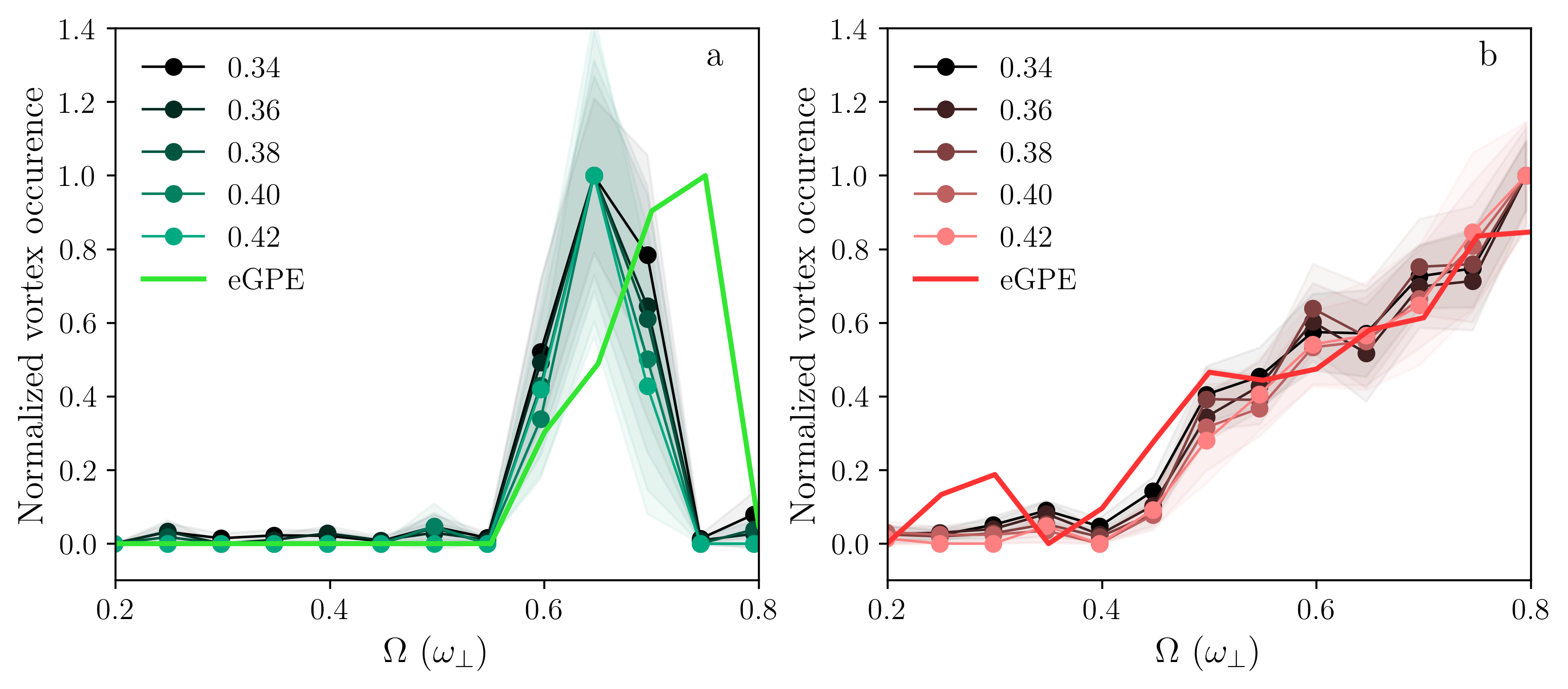}
    \caption{\textbf{Experimental vortex detection as a function of the threshold parameter.} Normalized vortex occurrence integrated over $1\,\si{s}$ of rotation in the BEC phase (left) and in the supersolid phase (right) as a function of the rotation frequency, for varying contrast threshold between $0.34$ and $0.42$ (see Extended Data Fig.\,\ref{fig:extendedFigResiduals}). The shaded areas indicate the error on the mean, \textit{i.e.} the standard deviation divided by the square root of the number of points ($8$). The solid lines are visual help. The results of the extended-GPE simulations (see Fig.\,\ref{fig:Fig3}) are plotted in thick solid lines as a comparison.}
    \label{fig:extendedFigIntegrated}
\end{figure}

\begin{figure}[t]
\centering    
\includegraphics[width=\textwidth]{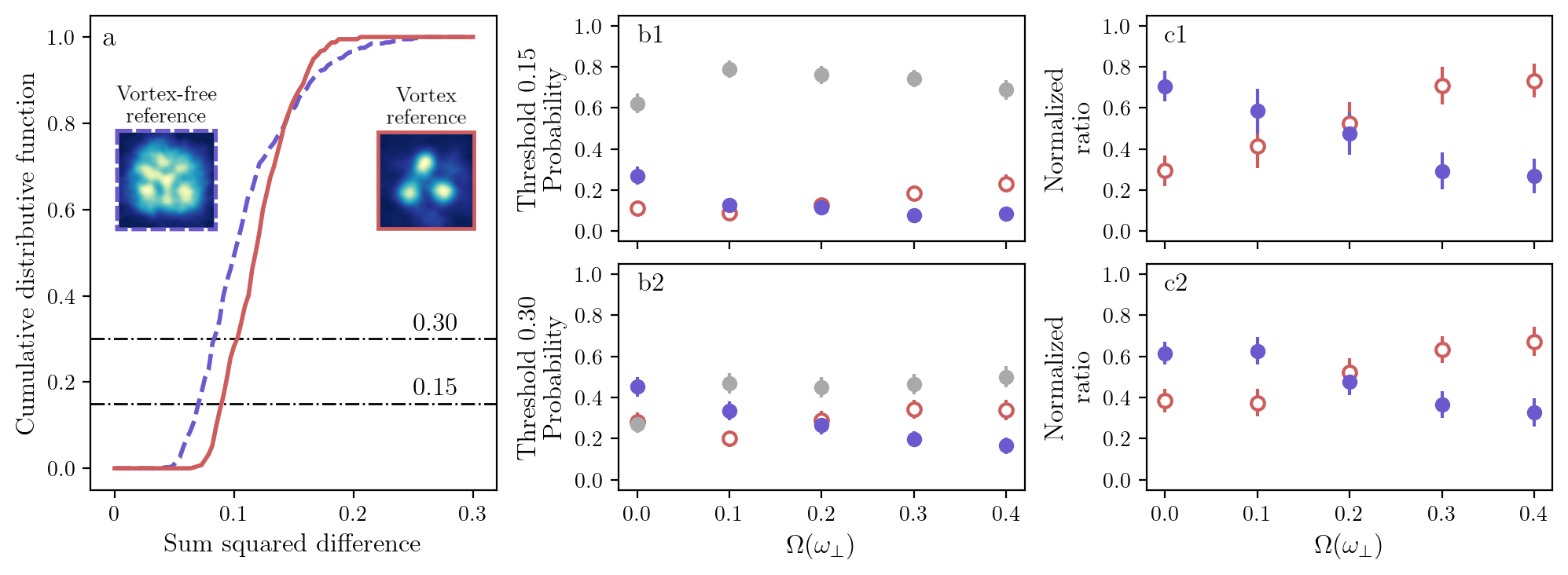}
    \caption{\textbf{Probability of detecting a vortex as a function of the rotation frequency}. \textbf{a} Cumulative distribution function obtained from the calculated sum squared differences over the whole data set, with each of vortex (solid line) and vortex-free (dashed line) references (see inset images). \textbf{b} With a defined threshold $X$ (dashed-dotted lines on \textbf{a}) on the cumulative distribution function, each image is assigned to a category: vortex (red empty circles), vortex-free (blue filled circles), or no classification (grey filled circles). \textbf{c} Probability of detecting a vortex signal and vortex-free signal out of the selected images in \textbf{b}. The error bars indicate the Clopper-Pearson uncertainty associated with image classification. Top and bottom rows show the classification result for respective thresholds 0.15 and 0.30 on the cumulative distribution function, showing the independence of the signal from the threshold.}
    \label{fig:extendedFigTOF}
\end{figure}

Since in the experiment we do not have access to the phase, to have a better comparison, we have also applied the vortex detection algorithm described in the previous section to the simulated states. With this aim, we linearly ramp the states of Fig.\,\ref{fig:Fig3}b from $a_s = 93a_0$ to $104a_0$ over $1\,\si{ms}$ to melt the droplets and perform the short $3\,\si{ms}$ TOF for each time and frequency. We apply a Gaussian filter of different sizes $\sigma$ to match the experimental resolution, and implemented the vortex detection algorithm detailed above to the resulting images. The same algorithm and filter have also been applied to the unmodulated BEC data of Fig.\,\ref{fig:Fig3}a. The results are shown in Extended Data Fig.\,\ref{fig:extendedFig_melted_phase_diagram}. Both methods give the same qualitative behavior as a function of the rotation frequency. Applying a Gaussian filter on the theoretical data reduces the vortex number to the experimental values (which are in general affected by the imaging noise). Critically, the resonant and threshold behaviors are robust to all the detection methods.

\vspace{0.1cm}
\noindent{\bf Time-of-flight vortex detection algorithm}

\noindent In the interference pattern, a striking difference between a single vortex and a vortex-free state is the absence or presence of a central density feature. This feature provides us with another fingerprint of vortices, thus allowing for binary classification of the experimental TOF images and extraction of the vortex occurrence probability as a function of $\Omega$. In the following paragraphs, our classification protocol is described.

First, we prepare all the images, $n_\mathrm{i}$, by denoising them with a Gaussian filter of size $\sigma=2$ px and by normalizing to the maximum density, $\mathrm{max}(n_\mathrm{i})\,{=}\,1$. 
Among those, we then select two reference images, one for each case: the presence ($n^{\mathrm{v}}_\mathrm{r}$) or absence ($n^{\emptyset}_\mathrm{r}$) of a vortex; see insets in Extended Data Fig.\,\ref{fig:extendedFigTOF}a. These will be used to classify all images.

Then, using ‘Powell’ minimization\cite{Powell1964aem}, we translate and rotate each image to best overlap with the references. To quantify the similarity of the images to each reference image, we calculate the sum squared differences, $S^\mathrm{\{v,\,\emptyset\}}$, between $n_\mathrm{i}$ and $n^{\{\mathrm{v},\,\emptyset\}}$.
Here, high values of $S^\mathrm{\{v,\,\emptyset\}}$ indicate large dissimilarity between the images.

We generate a cumulative distribution function for $S^\mathrm{v}$ and $S^\mathrm{\emptyset}$, which are normalized by the total number of images 
(see Extended Data Fig.\,\ref{fig:extendedFigTOF}a). 
Using the cumulative distribution, we generate one subset of images for each reference, which are the $X\%$ most similar images. The remaining images are not classified. Note that so far, the analysis is rotation frequency independent. Finally, we extract the number of images within each category as a function of rotation frequency ($\Omega$), see Extended Data Fig.\,\ref{fig:extendedFigTOF}b. 
Renormalizing to the total number of classified images, we obtain the ratio of images that have a central vortex, see Extended Data Fig.\,\ref{fig:extendedFigTOF}c.

At low rotation frequency, the vortex-free interference pattern is dominating. Crucially, the ratio of images with a vortex increases with increasing $\Omega$, consistent with our eGPE simulations and experimental findings shown in Fig.\,\ref{fig:Fig3}. This result is robust against choice of the classification threshold $X$ as shown in Extended Data Fig.\,\ref{fig:extendedFigTOF}c(1-2) for $X\,{=}\,15\,\%$ and $X\,{=}\,30\,\%$ (see dashed-dotted line in Extended Data Fig.\,\ref{fig:extendedFigTOF}a). Note that fluctuations of the experimental parameters lead to a non-zero vortex signal even without rotation. Note that the selection threshold is kept low, ensuring unambiguous categorization of the images.

\clearpage

\newpage
\onecolumngrid
\newpage
\clearpage
\newpage


\clearpage
\newpage

\clearpage

\newpage

\newpage

\clearpage

\newpage

\end{document}